\documentclass[11pt,amsmath,amssymb,aps,floatfix,preprintnumbers,nofootinbib,nobibnotes,bibnotes]{revtex4-2}

\usepackage{graphicx}
\usepackage{dcolumn}
\usepackage{bm}
\usepackage{hyperref}
\usepackage[mathlines]{lineno}
\usepackage{epsfig}  
\usepackage{color}
\usepackage{slashed}
\usepackage{float}

\def\bsll{b \to s \ell^+ \ell^-}

\def\bsmu{b \to s \mu^+ \mu^-}

\def\beq{\begin{equation}}
\def\eeq{\end{equation}}
\def\bea{\begin{eqnarray}}
\def\eea{\end{eqnarray}}

\def\Z{Z^{\prime}}

\large

\begin{document}

\title{A global analysis of $b \to s \ell \ell$ data in heavy and light $Z'$ models}

\author{Ashutosh Kumar Alok}
\email{akalok@iitj.ac.in}
\affiliation{Indian Institute of Technology Jodhpur, Jodhpur 342037, India}

\author{Neetu Raj Singh Chundawat}
\email{chundawat.1@iitj.ac.in}
\affiliation{Indian Institute of Technology Jodhpur, Jodhpur 342037, India}

\author{Shireen Gangal}
\email{gangal@lapth.cnrs.fr}
\affiliation{LAPTh, Universite Savoie Mont-Blanc et CNRS, Annecy, France}

\author{Dinesh Kumar}
\email{dinesh@uniraj.ac.in}
\affiliation{Department of Physics, University of Rajasthan, Jaipur 302004, India}

\begin{abstract}
We perform a model-independent global fit to all $b \to s \ell \ell$ data in the light of recent measurements of the lepton flavour universality violating (LFUV) observables $R_{K_S^0}$, $R_{K^{*+}}$, and the updated observables in $B_s \to \phi \mu^+ \mu^-$ decay, by the LHCb collaboration. 
Assuming NP only in the muon sector, we find that the 1D NP scenarios $C_9^{\rm NP} <0 $ and $C_{9}^{\rm NP}=-C_{10}^{\rm NP}$ continue to be the most favoured ones. However, the significance of the favoured scenario $C_{9}^{\rm NP}=-C'_{9}$ has reduced and the updated data now marginally prefers $C_{10}^{\rm NP} > 0$ scenario. 
The 2D scenarios $(C_9^{\rm NP}, C_{10}^{\prime} )$, $(C_9^{\rm NP}, C_{9}^{\prime} )$ and $(C_9^{\rm NP}, C_{10}^{\rm NP} )$, continue to be favoured by the data in the listed order. 
We analyse generic TeV scale $Z'$ models which can generate the 1D scenarios, $C_9^{\rm NP} <0 $ and
$C_9^{\rm NP} = -C_{10}^{\rm NP}$ as well as 2D scenarios $(C_9^{\rm NP}, C_{9}^{\prime} )$ and $(C_9^{\rm NP}, C_{10}^{\rm NP} )$. 
We find that all four models provide an equally good fit to the data. We also consider a model with a 25 MeV $Z'$ that couples to muons only and has a $q^2$ dependent $b - s$ coupling. 
We study the implications of the current data on the LFUV observables $R_{\phi}$, $Q_{4,5}$ as well as $R_{K^{(*)}}$ in the high $q^2$.  We find that the $Q_{4,5}$ observables have a potential to 
discriminate between a few favored  solutions, and disentangle different heavy and light $Z'$ scenarios.
\end{abstract}

\maketitle 

\newpage

\section{Introduction} 

Flavour physics, in particular decays involving $B$ mesons, is expected to play a pivotal role in probing physics beyond the Standard Model (SM) of electroweak interactions. 
Over the last decade, the currently running experiments at the LHC have provided a profusion of measurements in the neutral current $\bsll$ $(l=e,\,\mu)$ transitions, many of which
show intriguing discrepancies with the SM predictions. 
The most striking one is the discrepancy in the measurement of the ratio $R_K \equiv \Gamma(B^+ \to K^+ \mu^+ \mu^-)/\Gamma(B^+ \to K^+ e^+ e^-)$ which shows a  $3.1 \sigma$ deficit compared to the SM 
prediction in the (1.1 $\mathrm{GeV}^2 \le q^2 \le$ 6.0 $\mathrm{GeV}^2$) bin, where $q^2$ is the dilepton invariant mass-squared \cite{LHCb:2021trn}. Till date, this is considered to be the strongest evidence of lepton flavour universality violation (LFUV) in the $B$ sector. 
Further, the measurements of analogous ratio, $R_{K^*} \equiv \Gamma(B^0 \to K^{*0} \mu^+ \mu^-)/\Gamma(B^0 \to K^{*0} e^+ e^-)$, in the (0.045 $ \rm{GeV}^2 \le q^2 \le$ 1.1 $\rm{GeV}^2$) and 
(1.1 $\rm{GeV}^2 \le q^2 \le$ 6.0 $\rm{GeV}^2$) bins also disagree with the 
SM at $\sim$ 2.5$\sigma$ level \cite{rkstar}.  This has been reinforced 
by new LFU ratios recently measured by the LHCb collaboration in the channels, $B^0 \to K_S^0 \mu^+  \mu^- $ and $B^+ \to K^{*+} \mu^+ \mu^-$, which though consistent with the SM at $\sim 1.5 \sigma$ show a similar deficit  \cite{LHCb:2021lvy}. 
These measurements can be considered as invigorating signatures of new physics in $b \to s\, \mu^+ \mu^-$ or/and $b \to s\, e^+ e^-$ decays.
 
Apart from anomalous measurements of LFUV ratios, there are measurements of other observables in $B_s \to \phi\, \mu^+\,\mu^-$ and $B \to K^* \, \mu^+\,\mu^-$  decays which deviate from their SM predictions. In particular, 
the measured value of the branching ratio of $B_s \to \phi\, \mu^+\,\mu^-$ decay by LHCb collaboration exhibits tension with the SM at 3.5$\sigma$ level \cite{bsphilhc2,bsphilhc3}. 
Further, the measurement of the optimized angular observable  $P'_5$ in $B \to K^* \, \mu^+\,\mu^-$ decay by the LHCb in the (4.0 $\mathrm{GeV}^2 \le q^2 \le$ 6.0 $\mathrm{GeV}^2$) bin deviates from the SM prediction at the level of 3$\sigma$ \cite{Kstarlhcb1,Kstarlhcb2,LHCb:2020lmf,sm-angular}. The measured value of the branching ratio of the decay $B_s \to \mu^+ \mu^-$ by the LHCb, ATLAS and CMS collaborations also disagrees with the SM prediction at the level of 2$\sigma$ \cite{LHCb:2021awg,Combination,ATLAS:2018cur,CMS:2019bbr,LHCb:2017rmj,Altmannshofer:2021qrr}. These anomalies can be attributed to beyond SM contribution in $b \to s \mu^+ \mu^-$ process. 

The above anomalous measurements can be consistently analysed in a model-independent framework using the language of effective field theory \cite{Descotes-Genon:2013wba,Altmannshofer:2013foa,Alok:2017jgr,Alok:2019ufo,Altmannshofer:2021qrr,Carvunis:2021jga,Alguero:2021anc,Geng:2021nhg,Hurth:2021nsi,Angelescu:2021lln}. These analyses differ mainly in the treatment of hadronic uncertainties and the statistical approach. However, irrespective of the adopted methodology, new physics (NP) scenarios with non-zero muonic Wilson coefficients (WCs) corresponding to vector ($V$) and axial-vector ($A$) operators are statistically favoured. 
Among the possible $V$ and $A$ operators, we consider NP contribution to the operators ${\cal O}_9 = (\bar{s}\gamma_\mu P_L b)(\bar{\mu} \gamma^\mu \mu)$  and  ${\cal O}_{10} = (\bar{s}\gamma_\mu P_L b) (\bar{\mu} \gamma^\mu \gamma^5 \mu)$ already existing 
in the SM, and their chirally flipped counterparts ${\cal O}_9^\prime = (\bar{s}\gamma_\mu P_R b)(\bar{\mu} \gamma^\mu \mu)$ and ${\cal O}_{10}^\prime = (\bar{s}\gamma_\mu P_R b)(\bar{\mu} \gamma^\mu \gamma^5 \mu)$. 
The favored solutions obtained as a result of such model-independent analyses can be accomplished in several NP models. Models with an addition of a $Z'$ having non-universal couplings to leptons, see for e.g. \cite{Chang:2010zy,Buras:2013qja,Chang:2013hba,Altmannshofer:2014cfa,Crivellin:2015mga,Crivellin:2015lwa,Sierra:2015fma,Crivellin:2015era,Allanach:2015gkd,Boucenna:2016wpr,Boucenna:2016qad,Altmannshofer:2016jzy,Crivellin:2016ejn,Ko:2017yrd,Arcadi:2018tly,Darme:2018hqg,Singirala:2018mio,Hutauruk:2019crc,Baek:2019qte,Biswas:2019twf,Han:2019diw,Calibbi:2019lvs,Alok:2019xub,Alok:2021ydy,Crivellin:2020oup,Alok:2021pdh,Ray:2022bxg}, are considered to be one of the simplest extensions of the SM wherein these favored NP scenarios can be generated. Most of these models require a heavy $Z'$ with mass in the TeV range. However a few models with $Z'$ mass in the GeV \cite{Cline:2017lvv,Sala:2017ihs,Alok:2017sui,Bishara:2017pje,Darme:2020hpo,Darme:2021qzw,Crivellin:2022obd} or even MeV \cite{Datta:2017pfz,Alok:2017sui,Datta:2017ezo} range were also shown to account for the $\bsll$ anomalies. 

Recently, in October 2021 the LHCb collaboration presented new measurements of  $R_{K^0_S}$, $R_{K^{* +}}$ \cite{LHCb:2021lvy} and updated measurements for several $B_s \to \phi \mu^+\mu^-$  observables \cite{bsphilhc3,LHCb:2021xxq}.
One of the goal of this work is to study the impact of these  measurements on the currently favored NP scenarios in a model-independent way, following the same statistical approach as in our previous work \cite{Alok:2019ufo}. For the scenarios which provide a good fit to the data, we obtain predictions for the additional LFUV observables $R_{\phi}$, $R_{K^{(*)}}$ in the high $q^2$ bin as well as $Q_{4,5}$ observables \cite{Capdevila:2016ivx}, and analyze their potential in 
distinguishing between these favored NP solutions.  
In addition, we study the simplest heavy $Z'$ models which can generate the WC patterns that are favoured from the model-independent fits.  The $Z'$ in these
simplified models couples to both $\bar{s} b$ and $\mu^+ \mu^-$ at tree level and contributes to the operators   ${\cal O}_{9,10}$ and ${\cal O}_{9,10}^\prime$. Such a $Z'$ gives rise to scenarios with WC combinations: 
i) $C_9^{\rm NP}$ , ii)  $C_9^{\rm NP}  = -C_{10}^{\rm NP}$,   iii) $(C_9^{\rm NP}, C_{10}^{\rm NP} )$ and   (iv) $(C_9^{\rm NP}, C_{9}^{\prime} )$. While for the model-independent analyses only $b \to s\, \ell \ell$ data is used, in the case of $Z'$ models, additional relevant constraints from, for example $B$-$\bar{B}$ mixing, neutrino trident, are also taken into account. 
The goal of this study is to investigate 
the efficacy of different variations of the $Z'$ models to accommodate the entire $\bsll$ data. 
We also consider a 25 MeV $\Z$ with couplings only to muon as proposed in ref.~\cite{Datta:2017ezo}. 

The paper is organized as follows: In sec.~\ref{method}, we discuss the methodology of the global fit and list the measurements included in the fit. We collect the results of our global fit in sec.~\ref{mi}, considering scenarios with one 
non-zero NP WC and scenarios with two non-zero WCs, at a time. In sec.~\ref{zp} we discuss the various heavy $Z'$ models that can generate the NP scenarios favoured by the current data. In Sec.~\ref{lzp} we explore a model with a 
light $\sim 25$ MeV $Z'$ as a candidate to explain the $\bsll$ data. We summarize our findings in Sec~\ref{concl}. 

 \section{Methodology}
\label{method}
We begin by presenting the list of observables in $b \to s \ell \ell$ decays that exhibit deviations from the predictions of the SM. These are divided into two categories depending on their sensitivity to hadronic uncertainties: 
the LFUV ratios and other observables involving only $ b \to s \mu^+ \mu^-$ transition. The former being ratio observables are considered clean due to the cancellations of theory uncertainties in the SM, 
while the latter have larger uncertainties stemming from form factors 
and charm loop contributions. Below we list the observables included in our global analyses. 
\begin{itemize}
\item {\it $R_K$ and $R_{K^*}$:} The first measurement of $R_K$ was reported by the LHCb collaboration in 2014 \cite{LHCb:2014vgu}. The measured value  in $1.0\, {\rm GeV}^2 \le q^2 \le 6.0 \, {\rm GeV}^2$ bin deviated from the SM prediction of  $R_K^{\rm SM} = 1\pm 0.01$ \cite{Bordone:2016gaq} by 2.5$\sigma$. This measurement was updated in Moriond 2021 and the measured value $R_K^{\rm exp}=0.846^{+0.044}_{-0.041}$ deviates from the SM at the level of 3.1$\sigma$ \cite{LHCb:2021trn}. 

The LFUV in $b \to s \ell \ell$ was further corroborated by the measurement of $R_{K^*}$ in the two different $q^2$ bins by the LHCb collaboration in 2017:
\begin{equation}
R_{K^*}^{\rm exp} =
\left\{
\begin{array}{cc}
0.660^{+0.110}_{-0.070}~ \pm 0.024~,~~ & 0.045 \le q^2 \le 1.1 ~{\rm GeV}^2 ~, \\
0.685^{+0.113}_{-0.069}~ \pm 0.047~,~~ & 1.1 \le q^2 \le 6.0 ~{\rm GeV}^2 ~.
\end{array}
\right.
\end{equation}
The SM prediction of $R_{K^*}$ in $0.045 \le q^2 \le 1.1 ~{\rm GeV}^2$ bin is $\simeq 0.93$  \cite{Straub:2018kue} whereas it is $\sim 1$ in $1.1 \le q^2 \le 6.0 ~{\rm GeV}^2$ \cite{Hiller:2003js}. Therefore, the measured values differ from the SM prediction at the level of $\sim$ 2.5$\sigma$. Apart from these measurements, $R_{K^*}$ was also measured by the Belle Collaboration in both $B^0$  and  $B^+$ decays \cite{Belle:2019oag}. We include in our global fit
Belle $R_{K^*}$ measurements in the bins $0.045 \,{\rm GeV}^{2}< q^2 < 1.1\, {\rm GeV}^2, \, 1.1 \,{\rm GeV}^{2}< q^2 < 6.0\, {\rm GeV}^2, $ and $15.0\, {\rm GeV}^{2}< q^2 < 19.0\, {\rm GeV}^2$.

\item  {\it $R_{K_S^0}$ and $R_{K^{*+}}$:} In October 2021, LHCb presented the first measurement of the ratio $R_{K_S ^0}\equiv \Gamma(B^0 \to K_S^0 \mu^+ \mu^-)/\Gamma(B^0 \to K_S^0 e^+ e^-)$ and $R_{K^{*+}}\equiv \Gamma(B^+ \to K^{*+} \mu^+ \mu^-)/\Gamma(B^+ \to K^{*+} e^+ e^-)$ 
corresponding to a luminosity of 9${\rm fb^{-1}}$ recorded in  2011 (7 TeV), 2012 (8 TeV) as Run 1 and 2016 - 2018 (13 TeV) as Run 2. The measured value of  $R_{K_S ^0}^{\rm exp} =  0.66^{+0.20+0.02}_{-0.14 - 0.04}$ and 
$R_{K^{*+}}^{\rm exp} =  0.70^{+0.18+0.03}_{-0.13 - 0.04}$ 
in the region $1.1\, {\rm GeV}^2 \le q^2 \le 6.0 \, {\rm GeV}^2$ show some deficit
but is consistent with the SM at 1.5$\sigma$ \cite{LHCb:2021lvy}.

\item {\it Branching ratio of  $B_s \to \mu^+ \mu^-$:} The LHCb updated the experimental value of the branching ratio of $B_s \to \mu^+ \mu^-$ in Moriond 2021  \cite{LHCb:2021awg}. The measured value,  $ \left(3.09^{+0.46+0.15}_{-0.43-0.11} \right) \times 10^{-9}$,  is nearly the same as the previous world average \cite{Combination}  of the measurements performed by the ATLAS \cite{ATLAS:2018cur}, CMS \cite{CMS:2019bbr} and LHCb \cite{LHCb:2017rmj} collaborations. In our fit, we consider the updated world average $ B(B_s \to \mu^+ \mu^-)= ( 2.93 \pm 0.35 ) \times 10 ^{-9} $ which is 2.3$\sigma$ away from the SM prediction \cite{Altmannshofer:2021qrr}.

\item {\it Differential branching ratios:} We update in our fit the recently measured the differential branching fraction measurements of $B_s \to \phi \mu^+ \mu^-$ by LHCb
in various  $q^2$ intervals \cite{bsphilhc3}.  We also include differential branching ratios of $B^0 \to K^{*0} \mu^+ \mu^- $ \cite{LHCb:2016ykl,CDFupdate,Khachatryan:2015isa}, $B^{+} \to K^{*+}\mu^{+}\mu^{-}$, $B^{0}\to K^{0} \mu^{+}\mu^{-}$and  $B^{+}\rightarrow K^{+}\mu^{+}\mu^{-}$ \cite{Aaij:2014pli,CDFupdate} in different $q^2$ bins. Further, the branching fraction of inclusive decay mode $B \to X_{s}\mu^{+}\mu^{-}$ \cite{Lees:2013nxa} where $X_{s}$ is a hadron containing only one kaon is included in the fit in the low and high-$q^2$ bins. 

\item {\it Angular Observables in $B^0 \to K^{*0} \mu^+ \mu^-$:} We include in our fit the longitudinal polarisation fraction $F_L$, forward-backward asymmetry $A_{FB}$  and observables  $S_3$, $S_4$, $S_5$, $S_7$, $S_8$, $S_9$
in various intervals of $q^2$, as measured by the LHCb collaboration in 2020 \cite{LHCb:2020lmf}, along with their experimental correlations.  
We also include the angular observables $F_L$, $P_1$, $P_4 '$, $P_5 '$, $P_6 '$ and  $P_8 '$ measured by ATLAS \cite{ATLAS:2018gqc} and $P_1$, $P_5 '$ measured by CMS \cite{CMS:2017rzx}. 
The measurements of $F_L$ and  $A_{FB}$ by CDF and CMS collaborations are also included \cite{ CDFupdate,Khachatryan:2015isa}. 

\item {\it Angular observables in $B^+ \to K^{*+} \mu^+\mu^-$:} The full set of angular observables for this decay mode was determined for the first time by LHCb in 2020 \cite{LHCb:2020gog}. The measured values show deviations from the SM predictions similar to those in the angular observables of $B^0 \to K^{*0} \mu^+\mu^-$ decay. Here, we take into account the results for $F_L$ and $P_1 - P_8 '$ optimized angular observables, along with their experimental correlation \cite{LHCb:2020gog}. 

\item {\it Angular observables in $B_s \to \phi \mu^+\mu^-$:} We include the $CP$-averaged observables $F_L$, $S_3$, $S_4$ and $S_7$ as measured by the LHCb in 2021 with the available experimental correlations \cite{LHCb:2021xxq}.
\end{itemize}

The discrepancies between the measurements and SM predictions of these LFUV ratios and angular observables suggests the presence of physics beyond the SM, which can be analyzed in a model-independent way 
using an effective field theory approach. The most general effective Hamiltonian for $ b \to s \mu^+ \mu^-$ decays in the presence of NP of the form of $V$ and $A$ operators is given by, 
\begin{eqnarray}
  \mathcal{H}_{\mathrm{eff}}(b \rightarrow s \mu \mu) =
  \mathcal{H}^{\rm SM} + \mathcal{H}^{\rm VA}  \; ,
\label{heff}
\end{eqnarray}
where the SM effective Hamiltonian is 
\begin{align} \nonumber
  \mathcal{H}^{\rm SM} &= -\frac{ 4 G_F}{\sqrt{2} \pi} V_{ts}^* V_{tb}
  \bigg[ \sum_{i=1}^{6}C_i \mathcal{O}_i + C_8 {\mathcal O}_8 
    + C_7\frac{e}{16 \pi^2}[\overline{s} \sigma_{\alpha \beta}
      (m_s P_L + m_b P_R)b] F^{\alpha \beta}  \nonumber\\
    & + C_9^{\rm SM} \frac{\alpha_{\rm em}}{4 \pi} 
    (\overline{s} \gamma^{\alpha} P_L b)(\overline{\mu} \gamma_{\alpha} \mu) 
    + C_{10}^{\rm SM} \frac{\alpha_{\rm em}}{4 \pi}
    (\overline{s} \gamma^{\alpha} P_L b)(\overline{\mu} \gamma_{\alpha} \gamma_{5} \mu)
    \bigg] \; .
\end{align}
Here $V_{ij}$ are the elements of the Cabibbo-Kobayashi-Maskawa
(CKM) matrix. The short-distance contributions are encoded in the WCs $C_i$ of the four-fermi operators ${\cal O}_i$. The scale-dependence here is implicit, i.e. $C_i \equiv C_i(\mu)$
and ${\cal O}_i \equiv {\cal O}_i(\mu)$. 
The operators ${\cal O}_i$ ($i=1,...,6,8$) contribute
through the modifications  $C_7(\mu) \rightarrow C_7^{\mathrm{eff}}(\mu,q^2)$
and $C_9(\mu) \rightarrow C_9^{\mathrm{eff}}(\mu,q^2)$.
The NP effective Hamiltonian can be represented as
\begin{align} \nonumber
  \mathcal{H}^{\rm VA} &=
  -\frac{\alpha_{\rm em} G_F}{\sqrt{2} \pi} V_{ts}^* V_{tb}
  \bigg[C_9^{\rm NP} (\overline{s} \gamma^{\alpha} P_L b)
    (\overline{\mu} \gamma_{\alpha} \mu)  
    + C_{10 }^{\rm NP} (\overline{s} \gamma^{\alpha} P_L b)
    (\overline{\mu} \gamma_{\alpha} \gamma_{5} \mu) \nonumber\\& 
    + C_9^{\prime}(\overline{s} \gamma^{\alpha} P_R b)(\overline{\mu} \gamma_{\alpha} \mu)  
    + C_{10 }^{\prime} (\overline{s} \gamma^{\alpha} P_R b)
    (\overline{\mu} \gamma_{\alpha} \gamma_{5} \mu) \bigg]\,,
    \label{heff-va}
\end{align}
where NP contributes through a change in the short-distance WCs and the NP WCs are denoted by
$C_9^{\rm NP}, C_{10}^{\rm NP},C_9^{\prime}$ and $C_{10}^{\prime}$. These NP WCs are assumed to be real in our analysis. 
We focus on scenarios where i) one of the NP WCs is non-zero or two of them are linearly related (``1D"), and ii) two of the WCs are non-zero at a time (``2D") so that we get six possible pairs. 
In order to identify the Lorentz structure of NP favoured by the $ b \to s \ell \ell$ data, we perform global fits based on a $\chi^2$ function that depends on these NP WCs and use the CERN minimization
code {\tt MINUIT} \cite{James:1975dr}.
The  $\chi^2$ function is defined as
\begin{equation}
  \chi^2(C_i,C_j) = \big[\mathcal{O}_{\rm th}(C_i,C_j) -\mathcal{O}_{\rm exp}\big]^T \,
  \mathcal{C}^{-1} \, \big[\mathcal{O}_{\rm th}(C_i,C_j) -\mathcal{O}_{\rm exp} \big]\,.
  \label{chisq-bsmumu}
\end{equation} 

\begin{table*}[h!]
  \begin{center}
\begin{tabular}{|c||c|c||c|c|}
\hline\hline
Wilson Coefficient(s) & \multicolumn{2}{|c|}{{\tt Moriond 2021}} & \multicolumn{2}{|c|}{Updated fits} \\ \hline
  & Best fit value(s) & $\Delta \chi^2_{2021}$ & Best fit value(s) & $\Delta \chi^2_{2022}$ \\  
\hline
$C_i=0\,\,\rm (SM)$ & - & 0 & - &  0 \\ 
\hline \hline
1D Scenarios: & & & &\\\hline		
$C_9^{\rm NP}$ & $-1.01 \pm 0.15$ & 48.34  & $-0.98 \pm 0.15$  & 44.35  \\ 
\hline 
$C_{10}^{\rm NP}$ & $0.71 \pm 0.13$ & 37.87   & $0.65 \pm 0.12 $  & 33.63  \\ 
\hline 
$C_9^{'}$  & $-0.05 \pm 0.13$& 0.17 & $ -0.15\pm 0.12$  & 1.59   \\ 
\hline 
$C_{10}^{'}$  & $-0.06 \pm 0.10$ &0.28  & $-0.04 \pm 0.09 $  & 0.26   \\ 
\hline 
$C_9^{\rm NP} = C_{10}^{\rm NP}$  & $0.16 \pm 0.15$ &1.23 & $0.13 \pm 0.13 $   &  1.00   \\ 
\hline 
$C_9^{\rm NP} = -C_{10}^{\rm NP}$  & $-0.49 \pm 0.07$ & 49.43  & $-0.46 \pm 0.07 $  &  44.73   \\
\hline 
$C_9^{'} = C_{10}^{'}$  & $-0.15\pm 0.13$ & 1.28  & $-0.08 \pm 0.13$  & 0.37  \\ 
\hline 
$C_9^{'} = -C_{10}^{'}$  & $0.01 \pm 0.06$& 0.02  & $-0.05 \pm 0.05 $  & 0.82  \\ 
\hline 
$C_9^{\rm NP} = C_{9}^{'}$   & $-0.38 \pm 0.10$ & 18.33 & $-0.38 \pm 0.09 $  & 20.22  \\ 
\hline 
$C_9^{\rm NP} = - C_{9}^{'}$  & $-1.03 \pm 0.15$& 44.55  & $-0.85 \pm 0.15$  & 29.48  \\ 
\hline 
$C_{10}^{\rm NP} = C_{10}^{'}$   & $0.31 \pm 0.09$& 15.01  &  $0.32 \pm 0.08$ & 18.43  \\ 
\hline 
$C_{10}^{\rm NP} = - C_{10}^{'}$  & $0.34 \pm 0.08$& 20.32  & $0.26 \pm 0.07 $  & 12.64   \\ 
\hline 
$C_{9}^{\rm NP} = C_{10}^{'}$   & $-0.54 \pm 0.10$ & 29.89 & $-0.39 \pm 0.09 $  & 16.77   \\ 
\hline 
$C_{9}^{\rm NP} = - C_{10}^{'}$  & $-0.20 \pm 0.07$& 8.60   & $-0.22 \pm 0.06 $   & 11.96  \\ 
\hline 
$C_{10}^{\rm NP} = C_{9}^{'}$   & $0.52 \pm 0.10$ & 27.24 & $0.41 \pm 0.10$  & 18.24  \\ 
\hline 
$C_{10}^{\rm NP} = - C_{9}^{'}$  & $0.32 \pm 0.08$ &19.07 & $0.32 \pm 0.07 $   & 20.61    \\ 
\hline \hline
2D Scenarios: & & & &\\ \hline
$(C_9^{\rm NP},C_{10}^{\rm NP})$  & (-0.82, 0.27) & 52.59  & $(-0.80, 0.24)$  & 47.94  \\  
\hline
$(C_9^{'},C_{10}^{'})$   & (-0.18, -0.14)& 1.38   & $(-0.22, -0.07) $  &  1.87    \\ 
\hline 
$(C_9^{\rm NP},C_{9}^{'})$   & (-1.19, 0.59) & 58.24  & $(-1.12, 0.40)$ & 49.54  \\  
\hline
$(C_9^{\rm NP},C_{10}^{'})$  &  (-1.26, -0.40) & 63.86   & $(-1.15, -0.26)$  & 51.51   \\ 
\hline 
$(C_{10}^{\rm NP},C_{9}^{'})$  &  (0.80, 0.24) & 40.73  & $(0.69, 0.10)$  & 34.27    \\  
\hline
$(C_{10}^{\rm NP},C_{10}^{'})$ &  (0.71, -0.04) & 38.02  & $(0.65, 0.04)$ & 33.79   \\  
\hline\hline
\end{tabular}
\caption{The best fit values of new WCs in various 1D and 2D scenarios. Here $\Delta\chi^2 = \chi^2_{\rm SM}-\chi^2_{\rm bf}$ where $\chi^2_{\rm bf}$ is the $\chi^2$ at the best fit point and $\chi^2_{\rm SM}$  corresponds to the SM. Note that $\chi^2_{\rm SM}  \approx$ 199 (200) for  Moriond 2021  (updated fits)  data set.}
\label{fit-1}
 \end{center}
\end{table*}

Here $\mathcal{O}_{\rm th}(C_i,C_j)$ are the theoretical predictions of the N=156 observables used in the fit, that depend on the NP WCs and
$\mathcal{O}_{\rm exp}$ are the corresponding central values of the experimental measurements. 
The $N \times N$ total covariance matrix $\mathcal{C}$ is obtained by adding the individual theoretical and experimental covariance matrices.
The theoretical covariance matrix includes uncertainties from input parameters, form factors and power corrections and is calculated using {\tt flavio} \cite{Straub:2018kue}.
The correlations among $\mathcal{O}_{\rm exp}$ are included for the angular observables in $B^0 \to K^{*0} \mu^+ \mu^-$ \cite{LHCb:2020lmf}, $B^+ \to K^{*+} \mu^+ \mu^-$ \cite{LHCb:2020gog}
and $B_s \to \phi \mu^+ \mu^-$ \cite{LHCb:2021xxq}. Wherever the errors are asymmetric,  we use the conservative approach of using the larger error on both sides of the central value.
The value of $\chi^2$ in the SM is denoted by $\chi^2_{\rm SM}$, and the best-fit value in the presence of NP WCs by $\chi^2_{\rm bf}$. 
We use $\Delta\chi^2 \equiv \chi^2_{\rm SM}-\chi^2_{\rm bf}$ for each NP scenario to quantify the extent to which a particular scenario is able to provide a better fit to the data compared to the SM.

\section{Analysis of $b \to s \ell \ell$ data: A model independent approach }
\label{mi}

\begin{figure*}[hbt]
\centering
\includegraphics[width = 2.1in]{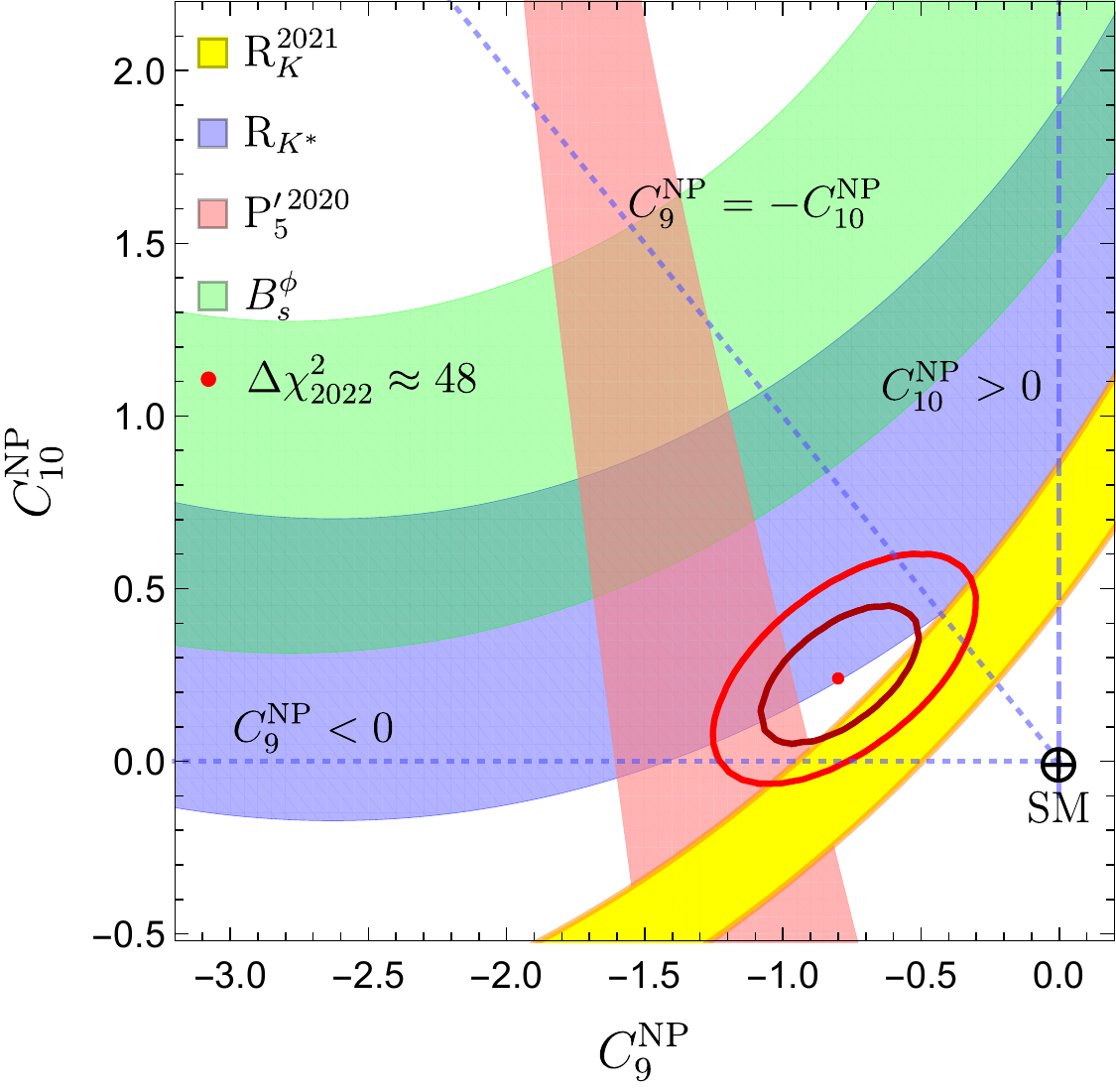}
\includegraphics[width = 2.1in]{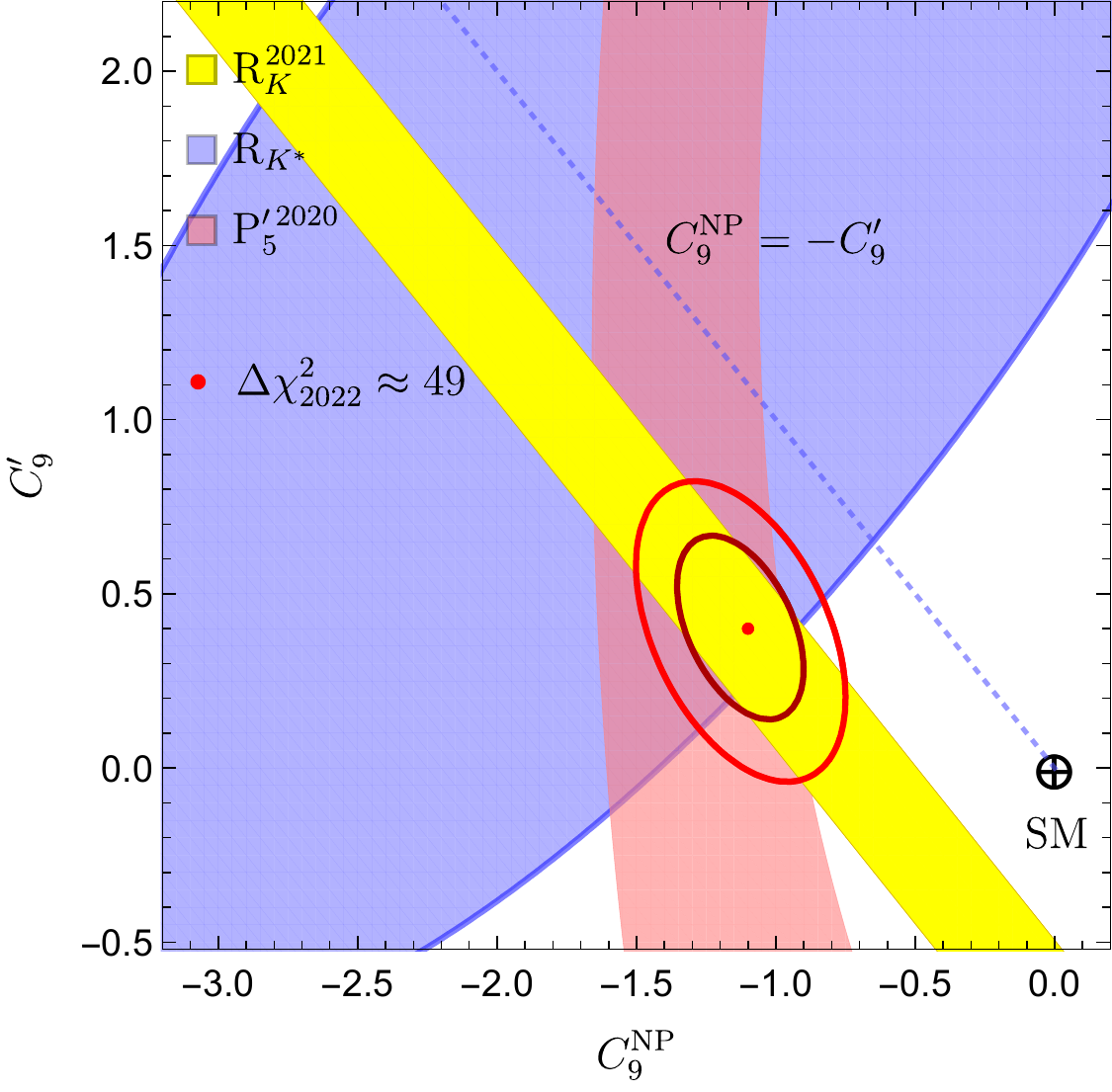}
\includegraphics[width = 2.1in]{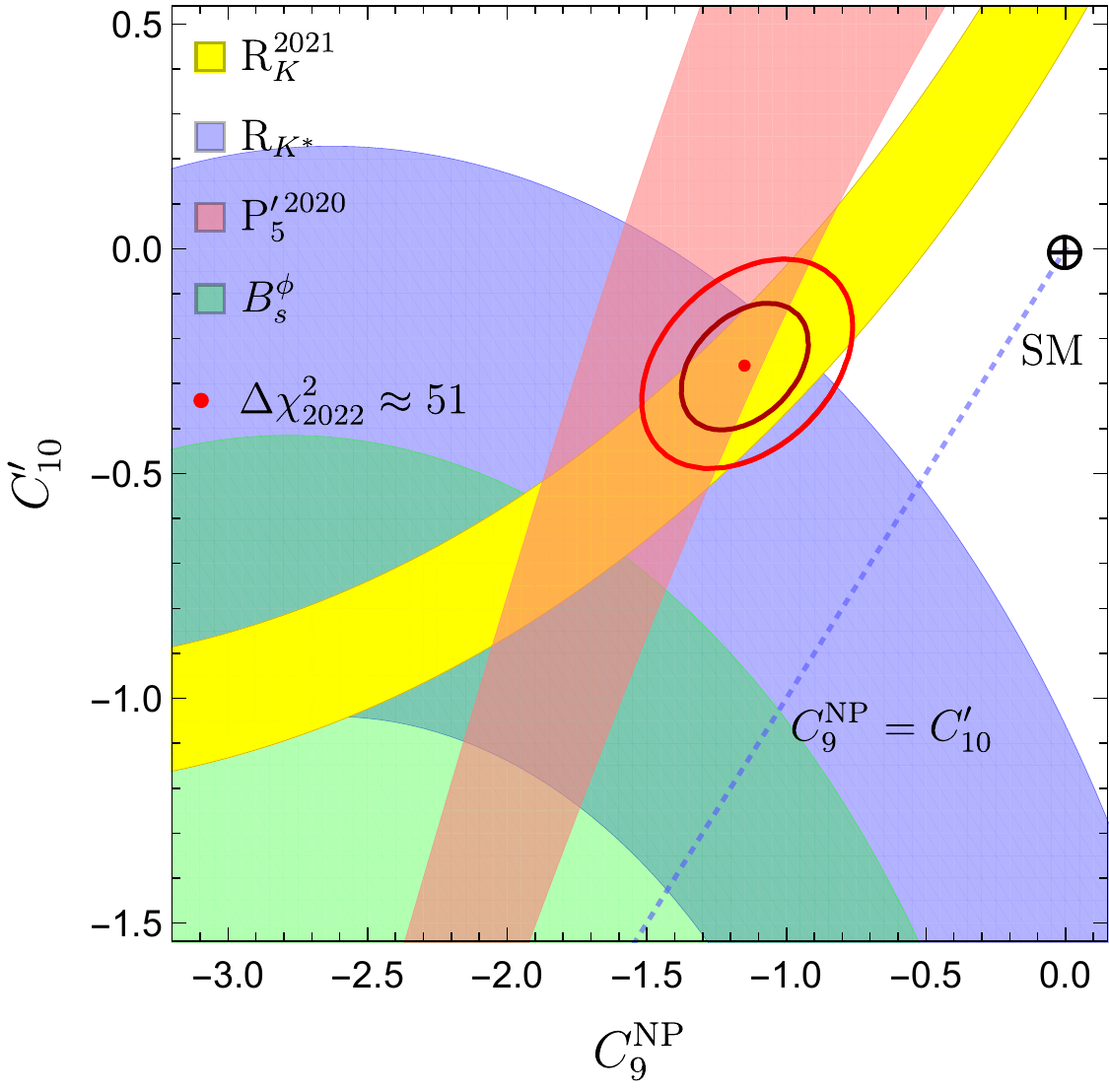}
\caption{Allowed parameter space for NP Scenarios  $(C_9^{\rm NP},C_{10}^{\rm NP})$, $(C_9^{\rm NP},C_{9}^{'})$  and  $(C_9^{\rm NP},C_{10}^{'})$.}
\label{fig:contour}
\end{figure*}

We consider ``1D" and ``2D" scenarios where NP affects the muon sector only.  In Table \ref{fit-1}, we update the results of ref.\cite{Alok:2019ufo}, in the following way: the fit 
results obtained after including the new measurements of $B_s \to \phi \mu \mu$, $R_{K^{*+}}$ and $R_{K^0_S}$  
by LHCb in 2021 are presented in the column ``Updated fits" $(\Delta\chi^2_{2022})$,  while the fit results obtained after Moriond 2021 are shown in 
``Moriond 2021" ($\Delta\chi^2_{2021}$). 
In fig.~\ref{fig:contour}, we show the $1\sigma$ allowed regions from the measurements of $R_K [1.1 - 6.0]$, 
$R_{K^{*}}[1.1-6.0]$, $P_5^\prime [4.0-6.0]$ and $B(B_s \to \phi \mu \mu)[1.0-6.0]$ with yellow, blue, pink and green bands
respectively. We superimpose on these bounds, the $1\sigma$ and $2\sigma$ contours obtained from our global fit to all
the $b \to s \ell \ell$ data. This allows us to check whether the best-fit regions are able to account for the above anomalies.

We find that the overall performance of the ``1D" scenarios in the updated fits after including the $B_s \to \phi \mu \mu$, $R_{K^{*+}}$ and $R_{K^0_S}$ measurements remain largely unchanged,
as compared to the one in ``Moriond 2021''.  
The scenarios: $C_9^{\rm NP} <0$ and $C_9^{\rm NP } = -C_{10}^{\rm NP}$ continue to provide a good fit to the data. It is evident from fig.~\ref{fig:contour} 
that these 1D scenarios along with 
$C_{10}^{\rm NP} >0$ scenario can alleviate the existing tension between the SM and the measured values of  $R_K$ and $R_{K^*}$ in the low-$q^2$ bin. The significance of the scenario  $C_9^{\rm NP} = -C_9^{'}$ has reduced considerably
compared to ref.~\cite{Alok:2019ufo} ($\Delta\chi^2_{2021} = 44$), as it  predicts $R_K \approx 1$, and hence fails to accommodate its current measurement. However, this scenario can accommodate the measurement of $R_{K^*}$ in the low-$q^2$ bin. The 
scenarios  $C_9^{\rm NP} <0$ and $C_9^{\rm NP} = -C_9^{'}$ can account for the measurement of $P'_5$ within 1$\sigma$, whereas the scenarios $C_9^{\rm NP} = -C_{10}^{\rm NP}$ and $C_{10}^{\rm NP} > 0$  can only provide a 
marginal resolution.
Further, none of the 1D and 2D scenarios can account for the anomalous measurement of the branching ratio of $B_s \to \phi\, \mu^+\,\mu^-$ decay, they can only invoke a marginal reduction in the existing tension.

In case of the six ``2D" scenarios, the overall conclusions remain the same as ref.~\cite{Alok:2019ufo}, though there is a slight reduction in the significance of the favourable scenarios: $(C_9^{\rm NP} , C_{10}^{\rm NP})$    
and $(C_9^{\rm NP} , C_{10}^{'})$. The three favoured scenarios  $(C_9^{\rm NP}, C_{10}^{\rm NP})$, $(C_9^{\rm NP} , C_{10}^{'})$ and $(C_9^{\rm NP} , C_{9}^{'})$ can explain $R_K$, $R_{K^*}$ and $P_5^\prime$ within 1$\sigma$. 

\begin{figure*}[hbt]
\centering
\includegraphics[width = 3.2in]{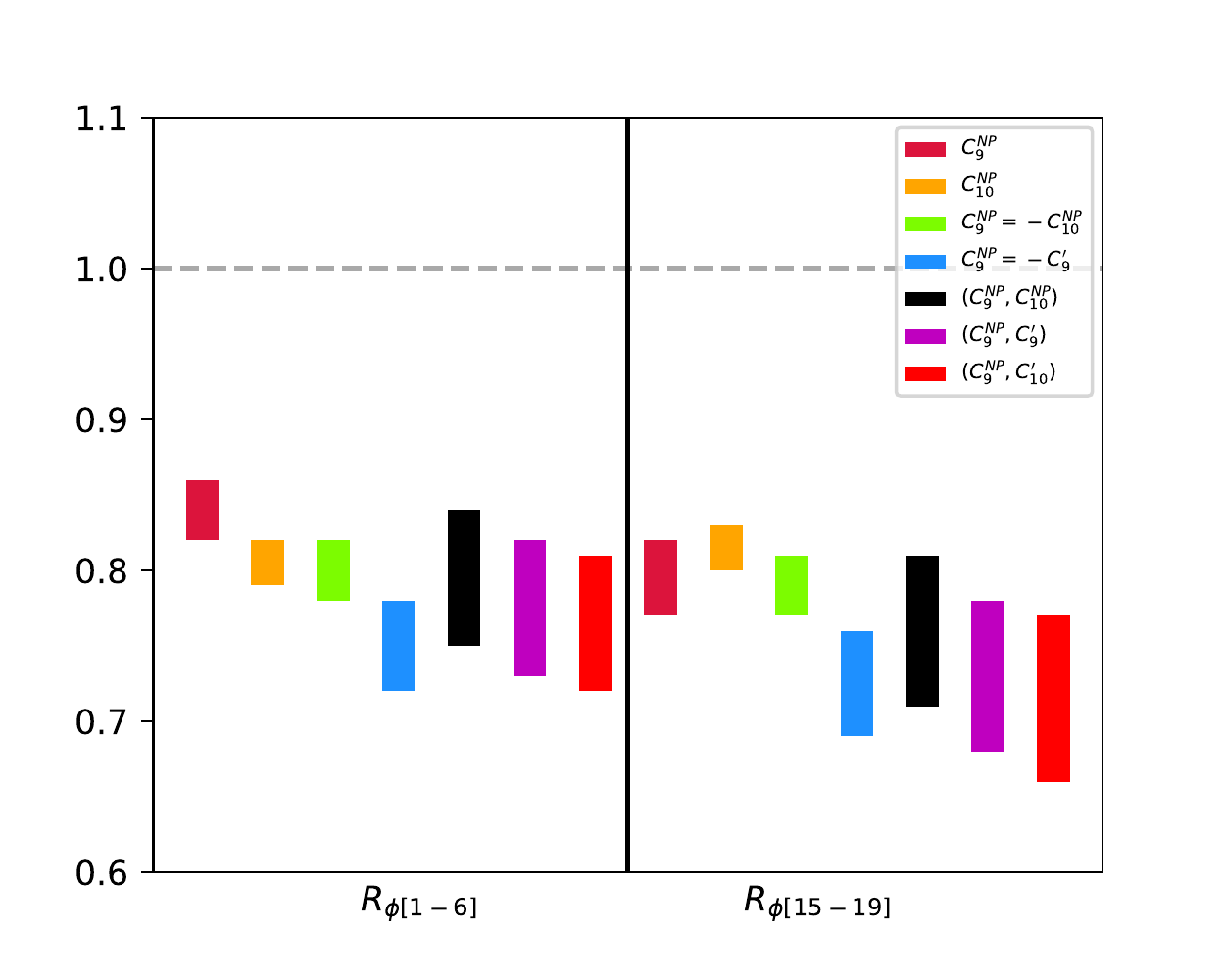}
\includegraphics[width = 3.2in]{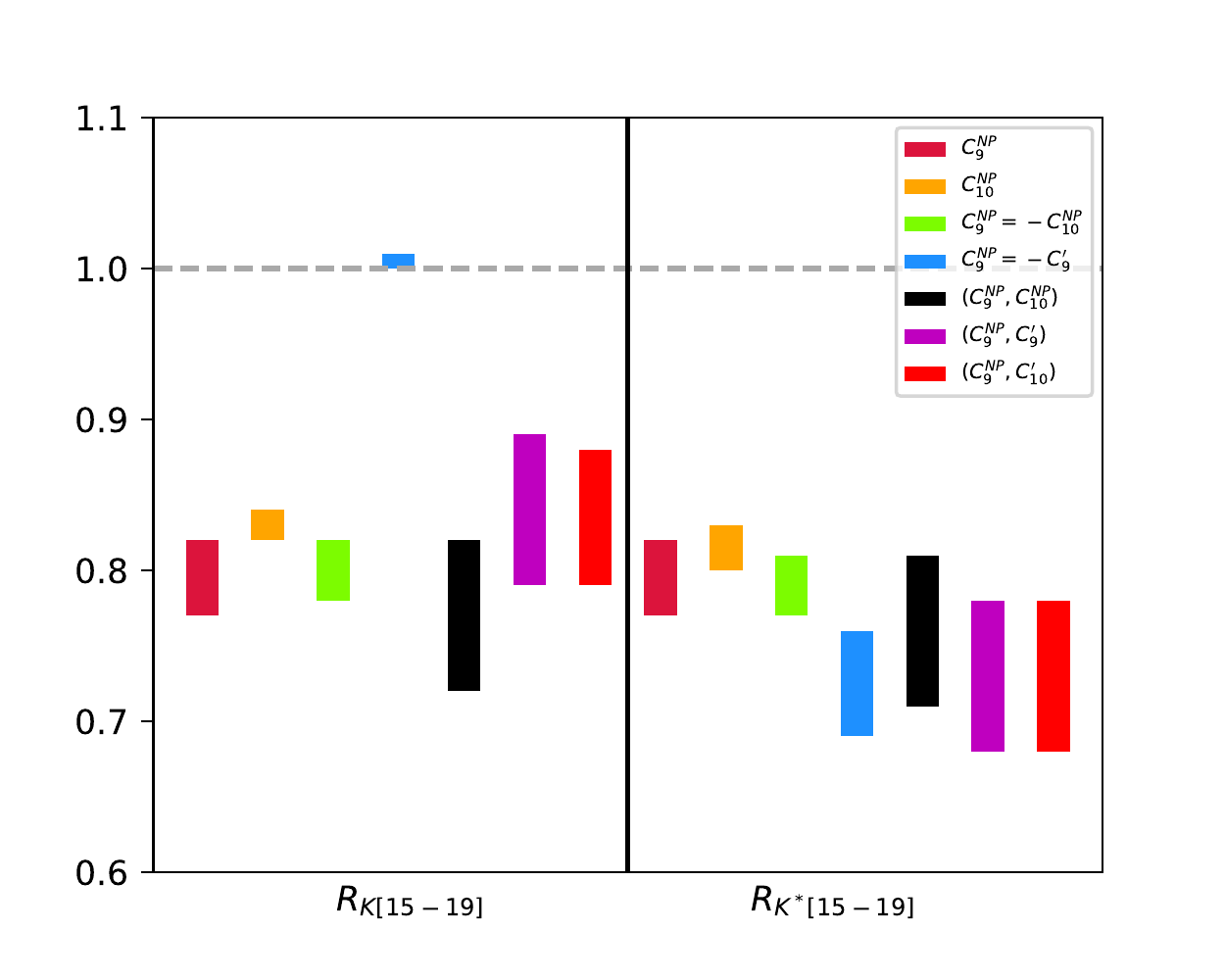}\\
\includegraphics[width = 3.2in]{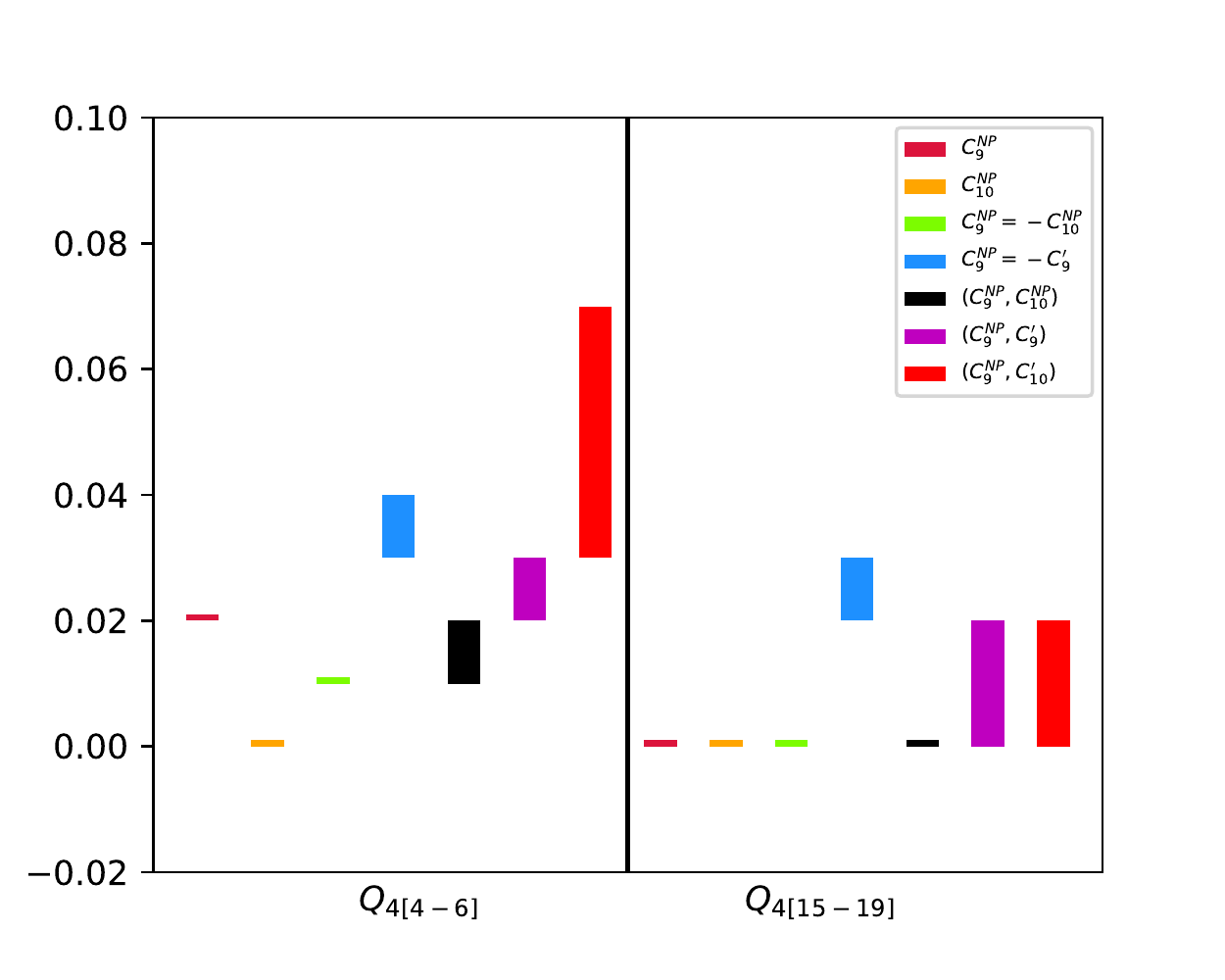}
\includegraphics[width = 3.2in]{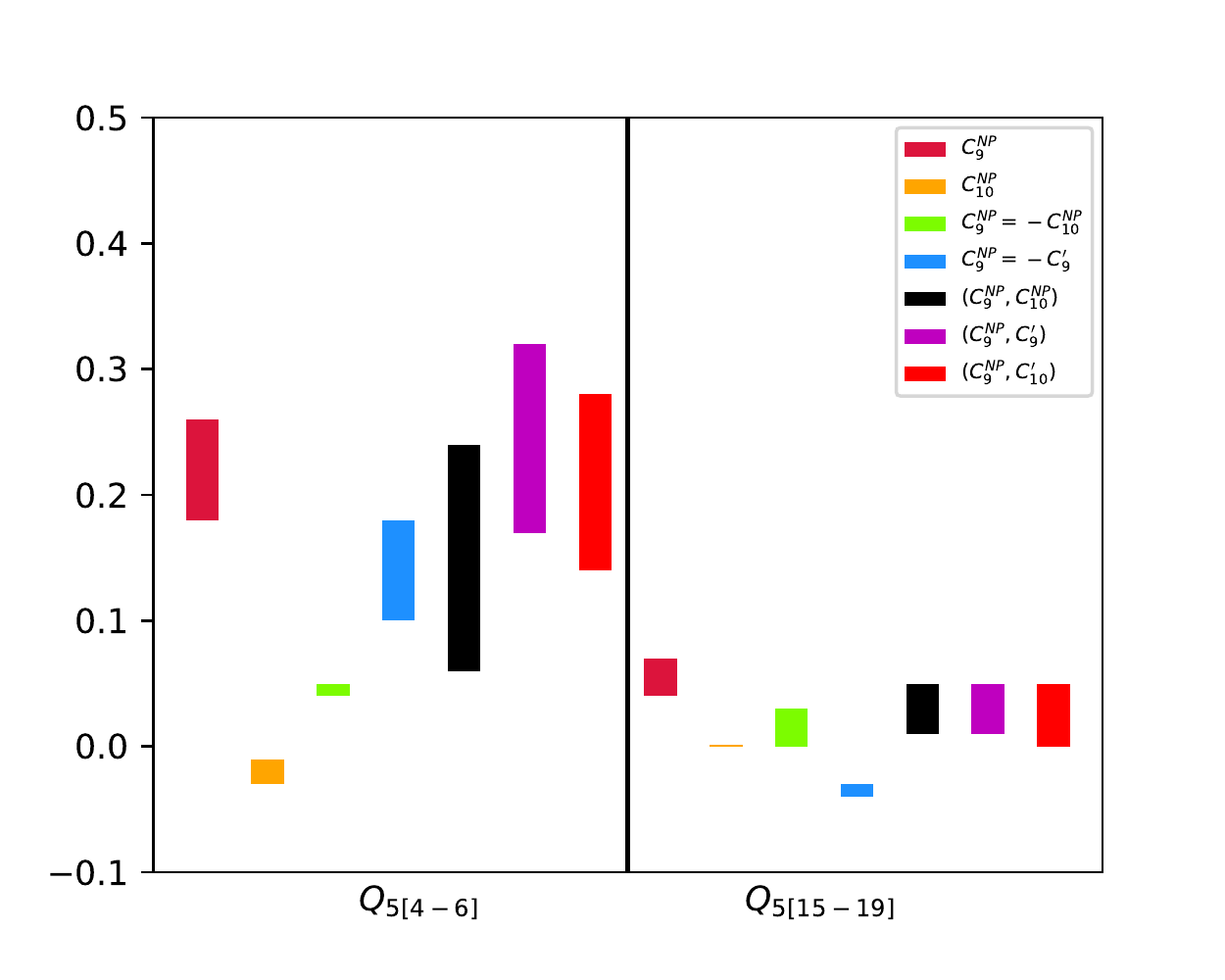}
\caption {Prediction (1$\sigma$ range) of additional LFUV observables for ``1D" and ``2D" solutions.}
\label{fig:pred-mi}
\end{figure*}

 The prediction of some of the additional LFUV observables are given in fig.~\ref{fig:pred-mi}. We consider the LFU ratios $R_{\phi}\equiv \Gamma(B_s \to \phi \mu^+ \mu^-)/\Gamma(B_s \to \phi  e^+ e^-)$ in both low and high-$q^2$ bins
and $R_{K^{(*)}}$ in the high $q^2$ bin.We also study the LFUV observable $Q_{4,5}$ which can be constructed as \cite{Capdevila:2016ivx}
 \begin{equation}
 Q_{4,5}=P^{'\mu}_{4,5} - P^{'e}_{4,5}\,.
 \end{equation}
 Within the SM, these observables are predicted to vanish to high accuracy. Hence any measurement of $Q$ observables different from zero would provide unambiguous signatures of NP. The $Q_4$ and $Q_5$ observables have been measured by the Belle collaboration \cite{Belle:2016fev}. However, owing to large errors, the measured values are consistent with zero.

All 1D scenarios predict $R_{\phi} <1$, both in the low and high-$q^2$ regions. The same is true for $R_{K^*}$ prediction in the high-$q^2$ bin. 
The scenario $C_9^{\rm NP} = - C_{9}^{'}$ is the only one that predicts $R_{K} \approx 1$ 
in the high-$q^2$ bin, and among the favored 1D scenarios, the only one that can invoke $R_\phi < 0.75$ and $R_{K^*} < 0.75$.  
Therefore this scenario can be distinguished from other 1D scenarios on the basis of future measurement of $R_{K}$ in the high $q^2$ bin. 
 
A large non-zero future measurement of ${Q_4}_{[15-19]}$ may indicate $C_9^{\rm NP} = - C_{9}^{'}$ as a more favourable scenario and the presence of chirally-flipped NP operators. 
The $C_9^{\rm NP} = - C_{9}^{'}$ scenario can also be distinguished from others based on the sign of ${Q_5}_{[15-19]}$ observable. 
Further, if ${Q_5}_{[4-6]}$ is measured with a value $>$ 0.20, then the $C_9^{\rm NP} < 0$ solution would be favoured over all the other 1D scenarios. 
A future measurement of ${Q_5}_{[4-6]} < 0$ may point towards the $C_{10}^{\rm NP}> 0$ scenario.

All 2D allowed scenarios predict $R_{\phi}<1$ as well as $R_{K^{(*)}}<1$ in the high-$q^2$ region.  
Hence, the values of $R_{\phi}$ in the low as well as high-$q^2$ bins do not have any kind of discrimination power. 
The  $(C_9^{\rm NP},C_{10}^{\rm NP})$ scenario is the only one which allows $R_K$ in the high-$q^2$ bin to be as low as $\sim$ 0.7.
While the $Q_{4_{[4-6]}}$ observable have similar $1\sigma$ allowed ranges for the 2D scenarios, larger values of this observable are allowed only in $(C_9^{\rm NP}, C_{10}^{'})$ scenario. 
Further, $Q_{4_{[15-16]}}$ is SM-like only for the $(C_9^{\rm NP},C_{10}^{\rm NP})$ scenario, and hence any large deviation could indicate the presence of chirally-flipped NP operators. 
Therefore any non-zero measurement  of ${Q_4}_{[15-19]}$ can distinguish this scenario from the other two.
The predicted values of $Q_{5_{[4-6]}}$ can be as large as $\sim 0.25$ for all 2D scenarios, and even larger in the case of $(C_9^{\rm NP}, C_9^{'})$ scenario.

 \section{Analysis of $b \to s \ell \ell$ data in 1D and 2D non-universal $Z'$ models}
\label{zp}

We consider a model with a $Z'$ boson in the TeV range that couples to $\bar{s}b$ as well as $\mu^+ \mu^-$.  This $Z'$ can be associated with a new $U(1)'$ symmetry. It couples to both left-handed and right-handed muons
but not to leptons of other generations. 
The $Z'$  couplings relevant for $b \to s \mu^+  \mu^-$  process can be written as \cite{Crivellin:2015era} 
\begin{equation}\nonumber
  \left[g^{bs}_L\, \bar{s} \gamma^{\alpha}P_L b   + g^{bs}_R\, \bar{s} \gamma^{\alpha}P_R b + g^{\mu\mu}_L\, \bar{\mu}\gamma^{\alpha}P_L\, \mu  + g^{\mu\mu}_R\, \bar{\mu}\gamma^{\alpha}P_R\, \mu 
\right]Z'_{\alpha} \,,
\label{eq:Jalpha}
\end{equation}
where $g_{L(R)}^{bs}$ are the left-handed (right-handed) couplings of the $Z'$ boson to quarks, and $g_{L(R)}^{\mu\mu}$ to muons. After integrating out  the heavy $Z'$ at the tree level, we get  the  effective four-fermion Hamiltonian which apart from contributing to $b\rightarrow s\, \mu^+\,\mu^-$ transition, also induces  $B_s - \bar{B_s}$ mixing. The relevant terms in the effective Hamiltonian are given by
\begin{eqnarray}
  \mathcal{H}_{\rm eff}^{Z'} &\supset &  \frac{g^{bs}_L}{M^2_{Z'}} \left(\bar{s}\gamma^{\alpha}P_L b\right)
  \left[\bar{\mu}\gamma_{\alpha}\left(g^{\mu\mu}_L P_L 
   + g^{\mu\mu}_R P_R\right)\mu \right] \nonumber \\ 
   &+& \frac{g^{bs}_R}{M^2_{Z'}} \left(\bar{s}\gamma^{\alpha}P_R b\right)
  \left[\bar{\mu}\gamma_{\alpha}\left(g^{\mu\mu}_L P_L
   + g^{\mu\mu}_R P_R\right)\mu \right] \nonumber \\
 & +& 
     \frac{\left(g^{bs}_L\right)^2}{2M^2_{Z'}}\left(\bar{s}\gamma^{\alpha}P_L b\right)\left(\bar{s}\gamma_{\alpha}P_L b\right)  \nonumber \\
 &+&  \frac{\left(g^{bs}_R\right)^2}{2M^2_{Z'}}\left(\bar{s}\gamma^{\alpha}P_R b\right)\left(\bar{s}\gamma_{\alpha}P_R b\right) \nonumber \\
 & +&  \frac{\left(g^{bs}_L g^{bs}_R\right)}{M^2_{Z'}}\left(\bar{s}\gamma^{\alpha}P_L b\right)\left(\bar{s}\gamma_{\alpha}P_R b\right)\,.
 \label{Leff}
 \end{eqnarray}
 
  \begin{table*}[h!]
\begin{center}
\begin{tabular}{|l|c|c|}
  \hline\hline
Scenario	& Couplings & Wilson coefficients  \\
  \hline
Z-I: $C^{\rm NP}_9$  & $g_R^{bs}=0, \, g_L^{bs}=0.0028;\, g_L^{\mu \mu}=g_R^{\mu \mu}=-0.28$ & $C_{9}^{\rm NP} = - N_1 g_L^{bs}g_L^{\mu \mu}$  \\ 
\hline
Z-II: $ C^{\rm NP}_9=-C^{\rm NP}_{10}$      &  $g_R^{bs}=0,  \, g_L^{bs}=0.0028;\, g_R^{\mu \mu}=0, \, g_L^{\mu \mu}=-0.26$   & $C_{9}^{\rm NP} = - C^{\rm NP}_{10}=- (N_1/2)\, g_L^{bs}g_L^{\mu \mu}$  \\ 
\hline
Z-III: $(C^{\rm NP}_9, \, C^{\rm NP}_{10})$  & $g_R^{bs}=0$, $g_L^{bs} = 0.002 $ &  $C_{9}^{\rm NP} = - (N_1/2)\,g_L^{bs}(g_L^{\mu \mu}+g_R^{\mu \mu})$  \\ 
& $g_L^{\mu \mu}=-0.29$,  $g_R^{\mu \mu} = -0.16$  & $C_{10}^{\rm NP} =  (N_1/2)\,g_L^{bs}(g_L^{\mu \mu}-g_R^{\mu \mu})$ \\
\hline
Z-IV: $(C^{\rm NP}_9,\, C_{9}^{\prime})$  & $g_R^{bs}=0.0005$, $g_L^{bs} = -0.001 $  &  $C^{\rm NP}_9 = -N_1 g_L^{bs}g_L^{\mu\mu}$  \\ 
& $g_L^{\mu \mu}=g_R^{\mu \mu}= 0.6$  & $C_{9}^{'} = - N_1 g_R^{bs}g_L^{\mu\mu}$ \\
\hline
  \hline
\end{tabular}
\caption{1D and 2D favoured NP scenarios that can be generated in $Z'$ model. }
\label{scenarios-heavyZ}
\end{center}
\end{table*}
 
The first two terms in eq.~(\ref{Leff}) induce $b\rightarrow s  \mu^+\mu^-$ transition.  These contributions modify the SM WCs $C^{\rm SM}_{9,10}$  as $C_{9,10} \rightarrow C^{\rm SM}_{9,10} + C^{\rm NP}_{9,10}$. The right-handed
quark couplings contribute to the chirally flipped WCs
$C_{9,10}^{\prime}$. Matching onto the effective Hamiltonian for $b\rightarrow s \mu^+\mu^-$ process as given in eq.~\eqref{heff-va}, the relevant WCs are, 
 \begin{eqnarray}
  C^{\rm NP}_9 &=& -(N_1/2)\, g_L^{bs}\,(g_L^{\mu\mu}+g_R^{\mu\mu})\,, \nonumber\\
  C^{\rm NP}_{10} &=& (N_1/2)\,  g_L^{bs}\,(g_L^{\mu\mu}- g_R^{\mu\mu})\,, \nonumber\\
  C_{9}^{\prime} &=& -(N_1/2)\,  g_R^{bs}\,(g_L^{\mu\mu}+g_R^{\mu\mu})\,, \nonumber\\
C_{10}^{\prime} &=& (N_1/2)\,  g_R^{bs}\,(g_L^{\mu\mu}- g_R^{\mu\mu})\,,
  \label{bsllNP}
\end{eqnarray}
where  $N_1 = \sqrt{2} \pi / (\alpha_{em} G_F V_{tb}V^*_{ts}M_{Z'}^2 )$. It is evident from eq.~\eqref{bsllNP} that the $Z'$ model can generate several favoured NP scenarios which are listed in Table~\ref{scenarios-heavyZ}, 
along with the corresponding best-fit values of the $Z'$ couplings obtained after a fit. We consider 
the favoured ``1D" scenarios $C_9^{\rm NP} $ and 
$C_9^{\rm NP} = -C_{10}^{\rm NP}$ which we denote by Z-I
and Z-II respectively, and the ``2D" scenarios $(C^{\rm NP}_9, \, C^{\rm NP}_{10})$  
and $(C_{9}^{\rm NP},\, C_{9}^{\prime})$ denoted as Z-III and Z-IV, respectively.  In principle, the 1D scenarios $C_{10}^{\rm NP}$ and $C_9^{\rm NP} = -C_{9}^{'}$ can also be generated in the $Z'$ framework. However, we don't consider them due to the fact that they can only provide a moderate fit to the data with $\Delta \chi^2 \approx 30$ whereas, as we will see below, $\Delta \chi^2$ for other four scenarios are $> 42$. It is also discernible from eq.~\eqref{bsllNP} that the 2D favoured scenario $(C_9^{\rm NP}, C_{10}^{'})$ cannot be generated in this framework.

The last three terms in eq.~\eqref{Leff} give rise to $B_{s}$--$\bar{B}_{s}$ mixing. The NP contribution to $B_{s}$--$\bar{B}_{s}$ mixing can be described by the following effective Hamiltonian,
\begin{eqnarray}
\mathcal{H}_{\rm eff}^{\Delta B = 2} & \supset &  \frac{4 G_F}{\sqrt{2}} \left(V_{tb} V^*_{ts} \right)^2 \Big[ 
C^{bs}_1 \left( \bar{s}\gamma^{\alpha}P_L b \right)^2 
+ C^{bs}_2 \left(\bar{s}\gamma^{\alpha}P_R b\right)^2 \nonumber\\
 &&+ C^{bs}_3 \left( \bar{s}\gamma^{\alpha}P_L b \right) \left( \bar{s}\gamma_{\alpha}P_R b \right) \Big]  \,.
\label{heffmix}
\end{eqnarray}
Here  scalar and tensor operators are neglected as they are disfavoured by the current $b \to s \ell \ell$ data.  In the presence of these operators, the contribution to $B_{s}$--$\bar{B}_{s}$ mixing  normalized to the SM is given by \cite{DiLuzio:2019jyq},
\begin{align}
\frac{\Delta M_s^\text{SM+NP}}{\Delta M_s^\text{SM}} = 
\Bigg| 1 &+ \frac{\eta^{6/23}}{R_\text{loop}^\text{SM}} 
\Bigg\{ C^{bs}_1 + C^{bs}_2 
- \frac{C^{bs}_3}{2\eta^{3/23}} \Bigg[\frac{B_5}{B_1} 
\left( \frac{M_{B_s}^2}{(m_b + m_s)^2} + \frac{3}{2} \right) \nonumber \\
&
+ \frac{B_4}{B_1} \left( \frac{M_{B_s}^2}{(m_b + m_s)^2} + \frac{1}{6} \right) 
\left(\eta^{-27/23} - 1\right) \Bigg] \Bigg\} \Bigg| \, .
\label{deltams}
\end{align}
Here $\eta = \alpha_s(\mu_\text{NP}) / \alpha_s(m_b)$ and the SM loop function is given by 
$R^\text{loop}_\text{SM}  = (1.310 \pm 0.010) \times 10^{-3}$ \cite{DiLuzio:2019jyq}.  The bag parameters $B_i$ are defined  in \cite{DiLuzio:2019jyq}. Matching the last three terms of eq.~\eqref{Leff} onto eq.~\eqref{heffmix}, the NP WCs at the scale $\mu = M_{Z^\prime}$ are given as
\begin{eqnarray}
C^{bs}_1 &=& \frac{1}{4 \sqrt{2} G_F M^2_{Z^\prime}} \left( \frac{g_L^{bs}}{V_{tb} V^*_{ts}} \right)^2 \,, \nonumber\\
C^{bs}_2 &=& \frac{1}{4 \sqrt{2} G_F M^2_{Z^\prime}} \left( \frac{g_R^{bs}}{V_{tb} V^*_{ts}} \right)^2 \,, \nonumber\\
C^{bs}_3 &=& \frac{\sqrt{2}}{4 G_F M^2_{Z^\prime}} \left( \frac{g_L^{bs} g_R^{bs}}{(V_{tb} V^*_{ts})^2} \right) \,.
\label{wcmix}
\end{eqnarray}
Using eq.~\eqref{deltams} and eq.~\eqref{wcmix}, we get
\begin{equation}
\frac{\Delta M_s^\text{SM+Z'}}{\Delta M_s^\text{SM}} \approx \left| 1 + 5 \times 10^{3}
\left\{ \left(g_L^{bs}\right)^2 + \left(g_R^{bs}\right)^2 - 9 g_L^{bs}g_R^{bs} \right\} \right| \,.
\label{mix-final}
\end{equation}

As the mass of  $Z'$ is assumed to be much above the electroweak scale,
the $Z'$ couplings must respect $SU(2)_L$ gauge invariance. Due to this symmetry,  the $Z'$ also couples to the left-handed neutrinos with $g^{\mu\mu}_L$ coupling. This induces an additional term in the effective Hamiltonian which can be written as 
\begin{equation}
  \mathcal{H}_{\rm eff}^{Z'} \supset  \frac{g^{\mu\mu}_L}{M^2_{Z'}} \left(\bar{\nu}_{\mu}\gamma_{\alpha}P_L\nu_{\mu}\right) \left[\bar{\mu}\gamma^{\alpha}\left(g^{\mu\mu}_L P_L  + g^{\mu\mu}_R P_R\right)\mu\right]\,.
\end{equation}
The above term  contributes to the neutrino trident production $\nu_{\mu} N \rightarrow \nu_{\mu} N \mu^+\mu^-$ ($N$ = nucleus) and modifies the cross section $\sigma$  as \cite{Alok:2017jgr}
\begin{eqnarray}
R_\nu =  \frac{\sigma}{\sigma_{\rm SM}} &=& \frac{1}{1+(1+4s^2_W)^2}\Bigg[\left(1+ \frac{v^2g^{\mu\mu}_L(g^{\mu\mu}_L-g^{\mu\mu}_R)}{M^2_{Z'}}\right)^2 \nonumber \\
&&
+ \left(1+4s^2_W+\frac{v^2g^{\mu\mu}_L(g^{\mu\mu}_L+g^{\mu\mu}_R)}{M^2_{Z'}}\right)^2\Bigg],\nonumber \\
\label{trident}
\end{eqnarray}
where $v=246$ GeV and $s_W = \sin\,\theta_W$.

We now consider contributions to the total $\chi^2$ coming from the above observables. For $b\to s \mu \mu$ observables, the $\chi^2$ is given by eq.~\eqref{chisq-bsmumu} with WCs expressed in terms of the $Z'$ couplings as given in eq.~\eqref{bsllNP}. Thus these observables provide constraints on the product of couplings $g_{L,R}^{bs} \, g_{L,R}^{\mu\mu}$. The constraint coming from the mass difference $\Delta M_s$ on $g_{L,R}^{bs}$ can be translated in the following form using eq.~\eqref{mix-final}
\begin{equation}
(g_L^{bs})^2 + (g_R^{bs} )^2 - 9\,g_L^{bs}\,g_R^{bs} = (7.69 \pm 12.94) \times 10^{-6}\,,
\end{equation}
where  we have used  $\Delta M_s^{\rm SM}/\Delta M_s^{\rm Exp}=1.04^{+0.04}_{-0.07}$ \cite{DiLuzio:2019jyq}. 
Therefore the contribution of $\Delta M_s$ to the $\chi^2$ can be written as
\begin{equation}
\chi^2_{\Delta M_s} = \left(\frac{\left((g_L^{bs})^2 + (g_R^{bs} )^2 - 9\,g_L^{bs}\,g_R^{bs}\right)- 7.69  \times 10^{-6}}{12.94 \times 10^{-6}}\right)^2\,.
\end{equation}
The contribution to the total $\chi^2$ coming from neutrino trident production, which constraints individual muon couplings $g_{L,R}^{\mu\mu}$, is given by
\begin{equation}
\chi^2_{\rm {trident}} = \left(\frac{R_\nu - 0.82}{0.28}\right)^2\,,
\end{equation}
where the theoretical expression of $R_\nu$ is given in eq.~(\ref{trident}), and the experimental value is $0.82 \pm 0.28$ \cite{Mishra:1991bv,Altmannshofer:2019zhy}. The total $\chi^2$ is given by 
\begin{equation}
\chi^2_{Z'_{\rm Heavy}} = \chi^2_{b\to s \mu \mu} + \chi^2_{\Delta M_s}   + \chi^2_{\rm trident}   \,.
\end{equation}

\begin{table}[hbt!]
\centering
\begin{tabular}{|c|c|c|c|c|c|c|c|} \hline
Scenario & Total $ \Delta \chi^2$ & \multicolumn{5}{|c|}{Prediction}   \\ 
\hline 
& &  ${R_K}_{[1.1-6]}$  & ${R_{K^*}}_{[0.045-1.1]}$ & ${R_{K^*}}_{[1.1-6]}$ & ${P'_{5}}_{[4-6]}$ & $B(B_s^{\phi \mu})_{[1-6]} \times 10^{-8}$ \\ \hline
Exp (1$\sigma$) & -- & [0.80, 0.89] & [0.57, 0.80]  & [0.55, 0.77]   & [-0.55, -0.32] & [2.67, 3.09] \\ \hline
Z-I & 42.16 & [0.75, 0.84]    & [0.89, 0.90]   & [0.82, 0.87]  & [-0.59, -0.46]    & [4.39, 4.70]  \\ \hline
Z-II &42.50 & [0.73, 0.83]   & [0.85, 0.88]  & [0.74, 0.83]   & [-0.72, -0.69]  & [4.01, 4.51]    \\ \hline
Z-III  & 45.61 & [0.71, 0.83]    & [0.86, 0.89]  & [0.74, 0.85]   & [-0.71, -0.48]  & [4.01, 4.60]  \\
\hline 
Z-IV  & 47.54 & [0.78, 0.90]   & [0.86, 0.89]   & [0.74, 0.84]  & [-0.59, -0.42]  &  [3.92, 4.48]  \\ \hline
\end{tabular}
\caption{Fit results for ``1D" and ``2D" scenarios generated in heavy $Z'$ models and 1$\sigma$ ranges of some of the observables used in the fit. Here $B_s^{\phi \mu}\equiv Br(B_s \to \phi \mu^+ \mu^-)$.}
\label{tab:2dzp}
\end{table}

\begin{figure}[H]
\centering
\includegraphics[width = 3.2in]{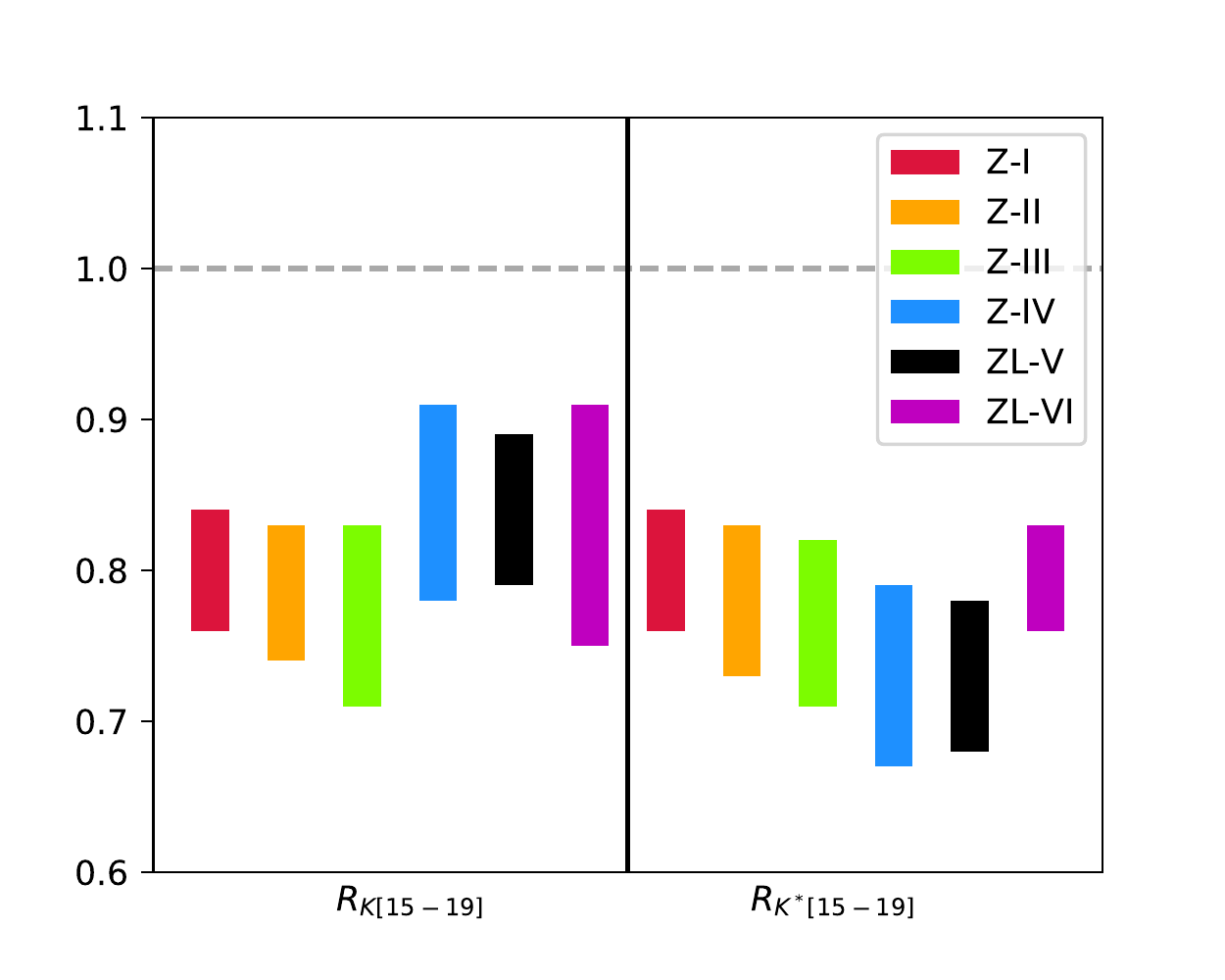}
\includegraphics[width = 3.2in]{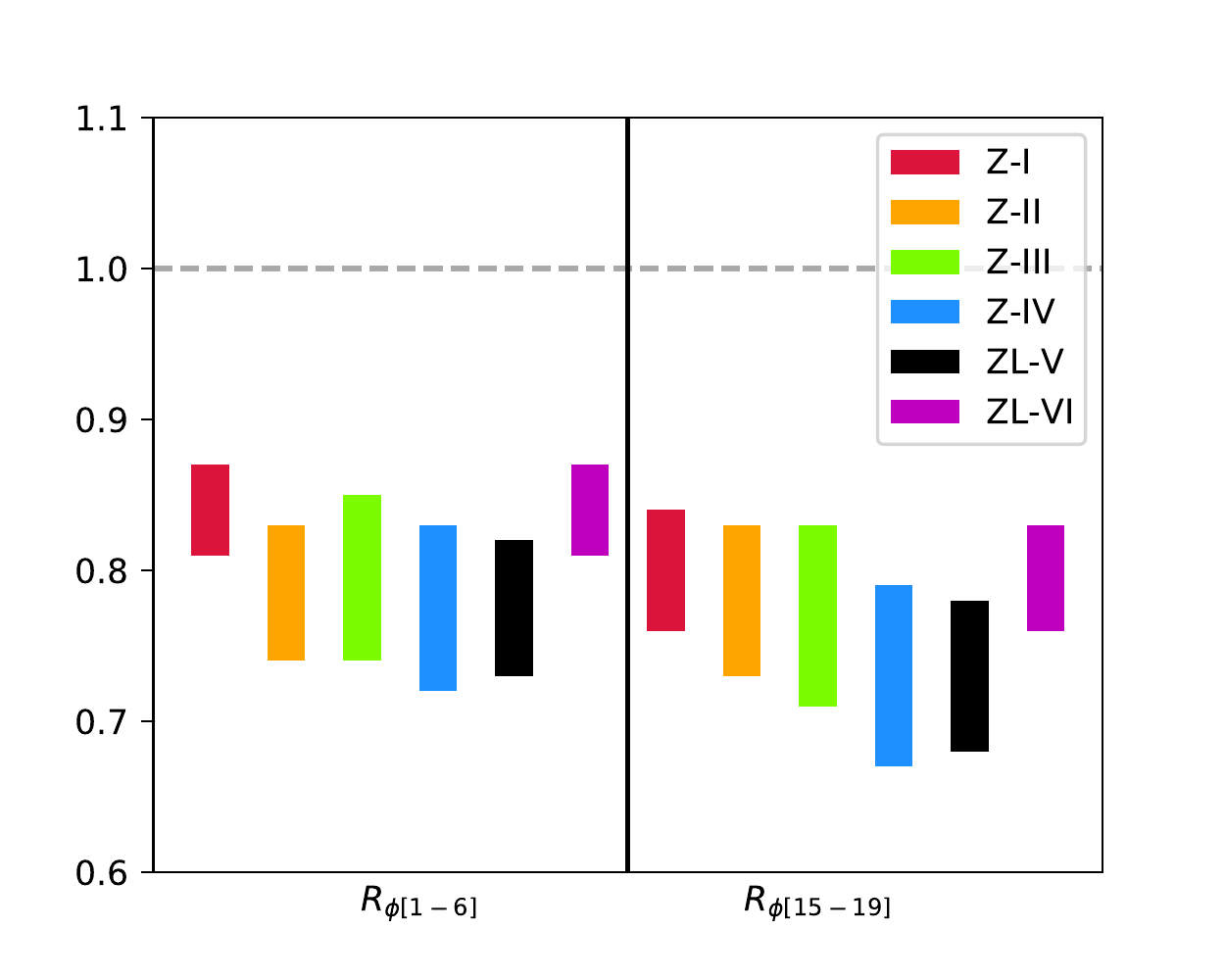}\\
\includegraphics[width = 3.2in]{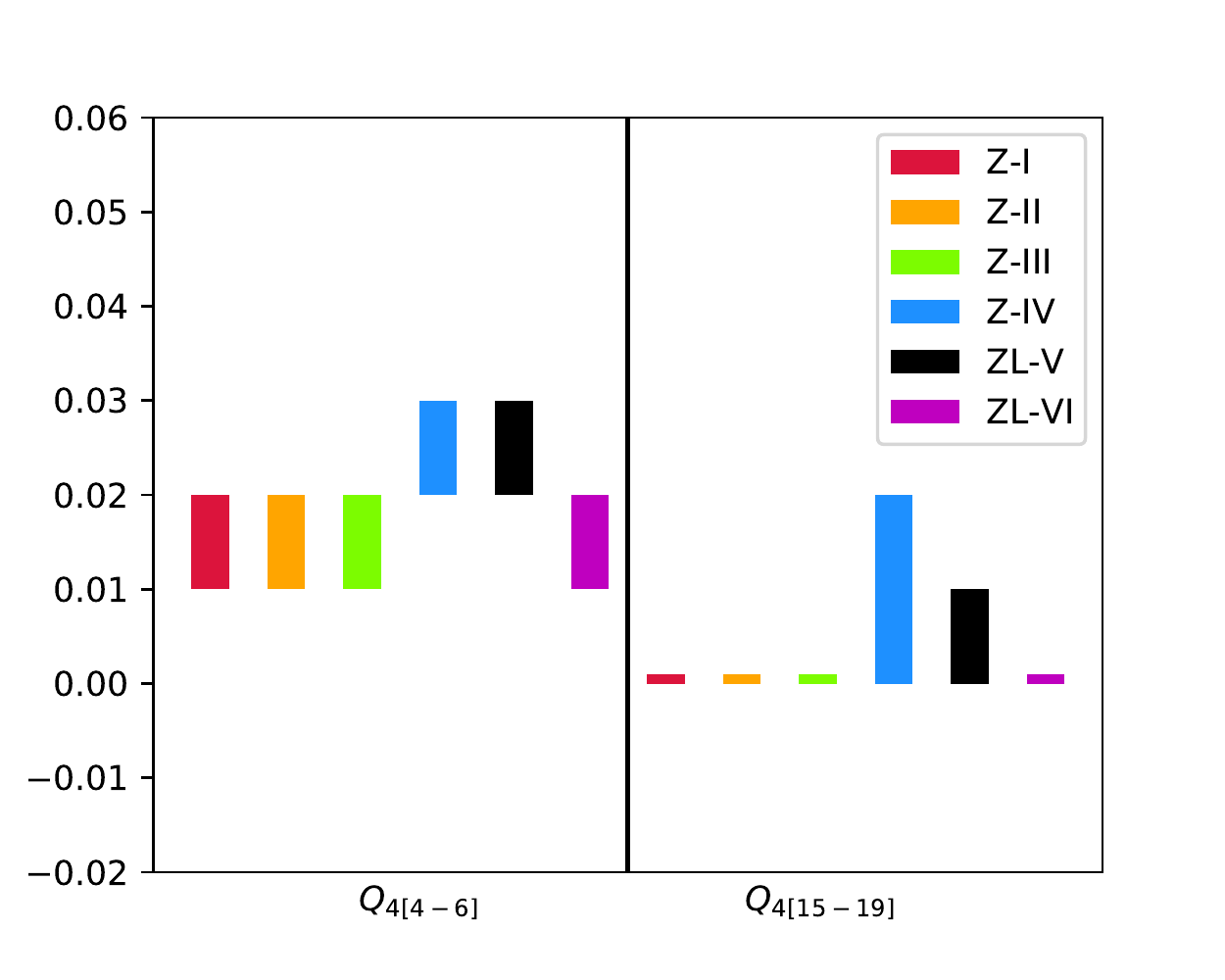}
\includegraphics[width = 3.2in]{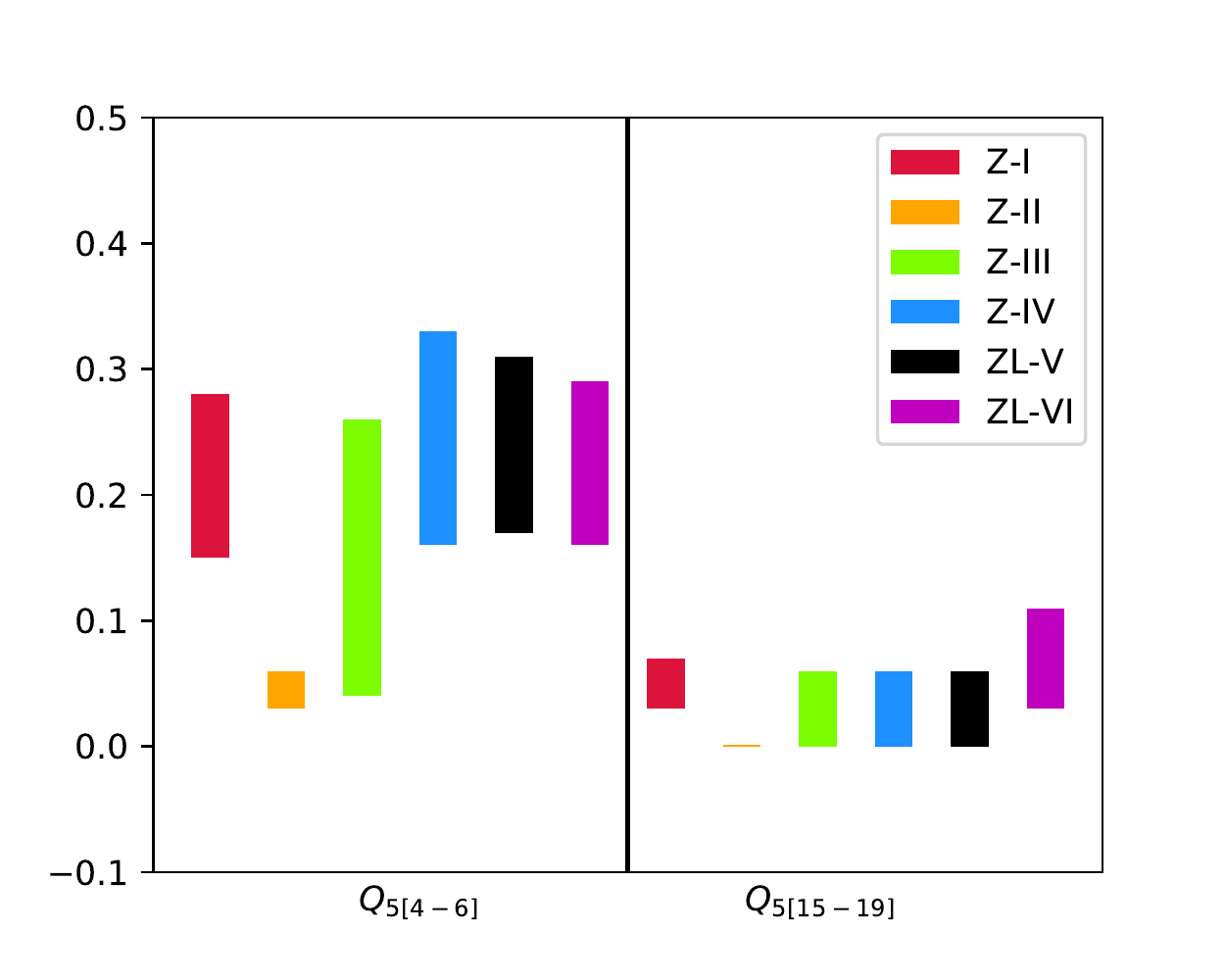}
\caption {Prediction  (1$\sigma$ range) of additional LFUV  observables in heavy and light $Z'$ models.}
\label{fig:psZprime1}
\end{figure}

Considering the mass of $Z'$ to be 1 TeV and assuming NP couplings to be real,
the fit results are given in Table \ref{tab:2dzp}. Following the methodology of the model independent analysis, the goodness of the fit is determined by $\Delta\chi^2 \equiv \chi^2_{\rm SM}-\chi^2_{\rm bf}$. The value of $\chi^2_{\rm SM}\approx$  200 and $\Delta\chi^2\approx  $ 42, 42, 46 and 48 for Z-I, Z-II, Z-III and Z-IV scenarios, respectively indicating that all the models provide a good fit to the data. The best fit values of $Z'$ couplings corresponding to these models are given in Table.~\ref{scenarios-heavyZ}. These scenarios can accommodate the measurement of ${R_{K^{(*)}}}$ in $[1.1-6.0]$ bin whereas only a marginal reduction in tension is possible for the branching ratio of $B_s \to \phi \mu^+ \mu^-$ in the low-$q^2$ bin.  The measurement of $P'_{5}$ in [4.0-6.0] bin can be explained within 1$\sigma$ for Z-I Z-III and Z-IV scenarios
whereas Z-II can only provide a minor improvement.

The 1$\sigma$ ranges of additional LFUV observables for various heavy $Z'$ models are shown in fig.~\ref{fig:psZprime1}. The ${R_{K}}$ values in [15-19] bin cannot provide any discrimination 
between models Z-I and Z-II and Z-III, though in Z-IV model the $R_{K_{[15-19]}}$ values can be close to $0.9$. The  ${R_{K^*}}_{[15-19]}$, $R_{\phi}$ and ${Q_4}_{[4-6]}$ observables do not have any discriminative capability. 
All scenarios, except Z-IV predict ${Q_4}_{[15-19]}$ values to be SM-like.

Interestingly, the allowed $1\sigma$ ranges of ${Q_5}_{[4-6]}$ for 1D models Z-I and Z-II, 
along with 2D models Z-III and Z-IV are distinct: ${Q_5}_{[4-6]}$ is restricted to be $<$ 0.06 for Z-II, whereas it can be $>$0.15 for Z-I. 
Moreover any measurement of ${Q_5}_{[4-6]}>0.25$ would favour Z-IV over Z-III. A precise measurement of $Q_5$ in [4-6] bin can therefore make it possible to distinguish between Z-I and Z-II as well as Z-III and Z-IV. 
The  ${Q_5}_{[15-19]}$ values cannot discriminate between models, however, any non-zero measurement would rule out Z-II scenario.

 In our analysis, we have assumed that the $Z'$ couples only to muons. This can be achieved by the $L_{\mu}$ symmetry. However this symmetry by itself is not anomaly-free and hence one requires additional symmetries. The gauged $L_{\mu}-L_{\tau}$ symmetry is one such popular choice \cite{Altmannshofer:2014cfa,Crivellin:2015mga,Crivellin:2015lwa,Altmannshofer:2016jzy,Crivellin:2016ejn,Ko:2017yrd,Arcadi:2018tly,Singirala:2018mio,Hutauruk:2019crc,Baek:2019qte,Biswas:2019twf,Han:2019diw,Crivellin:2020oup}. Due to this symmetry, new physics in muon sector will also generate effects in decays induced by the quark level transition $b \to s \tau^+ \tau^-$ \cite{Altmannshofer:2014cfa}.  Using the best fit values of NP couplings,   $B(B \to K \tau^+ \tau^-)_{[15-22]}$ and  $B(B \to K^* \tau^+ \tau^-)_{[15-19]}$ are predicted to be $1.34 \times 10^{-7} $ and $1.37 \times 10^{-7}$, respectively. These values are close to their respective SM predictions.

  \section{Analysis of $b \to s \ell \ell$ data a light $Z'$ model with $q^2$ dependent couplings}
\label{lzp}
  
One of the motivation for the light $Z'$ model is to explain the measurement of $R_{K^*}$ in the very low $q^2$ bin 0.045 $ \leq q^2 \leq$ 1.1 $\rm GeV^2$. In such models, the NP WCs have a $q^2$ dependence. This is due to the fact that the NP contributions cannot be integrated out. In ref.~\cite{Datta:2017pfz}, a $Z'$ with $q^2$ dependent $b-s$ couplings along with couplings to muon and neutrino, and with a mass less than two times the muon mass was proposed to explain the measurement of $R_K$ and the anomalous magnetic moment of the muon. A $Z'$ model with mass in a few GeV range was considered in ref.~\cite{Alok:2017sui}. However due to the mild dependence of WCs on $q^2$, only a marginal suppression in the value of $R_{K^*}$ in the very low-$q^2$ bin was possible.

In this work, we consider a 25 MeV $\Z$ model with couplings only to muons, first proposed in \cite{Datta:2017ezo}. 
Additionally,  a $\Z$ with couplings only to electrons were also introduced in ref.~\cite{Datta:2017ezo} in order to resolve the tension between the measured value of $R_{K^*}$ and the SM prediction in the very low-$q^2$ bin.
Our primary goal is to investigate the efficacy of different possible scenarios generated 
in this model to accommodate the entire $b \to s \mu^+ \mu^-$ data. This includes anomalous measurements of angular observable $P'_5$ in $B \to K^* \mu^+ \mu^-$  decay and the branching ratio of $B_s \to \phi \mu^+ \mu^-$. For the scenarios providing a good fit to the data, we obtain predictions for LFUV observables $R_{K^{(*)}}$ in the high-$q^2$ bin, $R_{\phi}$ and $Q_{4,5}$  to look for the possibility of distinguishing between the favoured scenarios
generated in this model.   
  
We assume the flavor-changing $bs\Z$ vertex to have the form,
\beq
F(q^2) \, \bar{s} \gamma^{\mu}\left[ g_{bs} P_L + g_{bs}' P_R \right]  b \, Z^\prime_{\mu} ~,
\label{bsZ}
\eeq
where for $q^2 \ll m_B^2$, the form factor $F(q^2)$ is defined as
\bea
F(q^2) & = & a_{bs} + b_{bs} \frac{q^2}{m_B^2} + \ldots\,.
\label{FF}
\eea

The matrix  elements for $b \to s \mu^+ \mu^-$  and the mass difference in $B_s$ mixing are
\bea
M_{b \to s \mu^+ \mu^-}
& = &
\frac{F(q^2)}{q^2- M_{Z'}^2} \left[\bar s \gamma^{\mu}\left( g_{bs} P_L + g_{bs}' P_R \right) b\right]
(\bar \mu \gamma^{\mu} (g_{L}^{\mu \mu}P_L + g_{R}^{\mu \mu}P_R) \mu) \nonumber\\
& - &
 \frac{ F(q^2)}{q^2- M_{Z'}^2} \frac{m_b m_{\mu}}{M_{\Z}^2}
 (g_{R}^{\mu \mu} - g_{L}^{\mu \mu}) 
\left[\bar s \left( g_{bs} P_R + g_{bs}' P_L \right) b \right]
(\bar \mu \gamma_5 \mu)\,, \\
\Delta M_{{B_s}}^{\rm NP} & = &
\frac{(F(q^2))^2}{2 q^2- 2 M_{Z'}^2} \frac{2}{3} f_B^2{m_{B_s}} \Bigg[ \left(g_{bs}^2+g_{bs}'^2 \right) \left(1- \frac{5}{8} \frac{m_b^2}{M_{\Z}^2} \right)  - 2 g_{bs}g_{bs}' \left(\frac{5}{6} - \frac{m_b^2}{M_{\Z}^2} \frac{7}{12} \right) \Bigg]\,.  
\label{Zp}
\eea
 Also, we define $g_{\ell\ell}\equiv(g_{L}^{\ell\ell} + g_{R}^{\ell\ell})/2$ and
 $g'_{\ell\ell}\equiv(g_{R}^{\ell\ell} -g_{L}^{\ell\ell})/2$ for convenience.
Including  the term proportional to $m_{\mu}$ in $M_{\bsmu}$, we get
 \begin{align}
 C_{9}^{\rm NP} &= (N_2/2) \left(a_{bs}g_{bs}+b_{bs}g_{bs}(q^2/m_{B}^2)\right) \left(g_R^{\mu \mu}+ g_L^{\mu \mu}\right)\,,\nonumber\\
  C_{10}^{\rm NP} &=  (N_2/2)  \left(a_{bs}g_{bs}+b_{bs}g_{bs}(q^2/m_{B}^2)\right) \left(g_R^{\mu \mu}- g_L^{\mu \mu}\right)\,,\nonumber\\
   C_{9}^{'} &= (N_2/2)  \left(a_{bs}g_{bs}^{'} +b_{bs}g_{bs}^{'} (q^2/m_{B}^2)\right) \left(g_R^{\mu \mu}+ g_L^{\mu \mu}\right)\,,\nonumber\\
   C_{10}^{'} &=  (N_2/2)  \left(a_{bs}g_{bs}^{'}+b_{bs}g_{bs}^{'}(q^2/m_{B}^2)\right) \left(g_R^{\mu \mu}- g_L^{\mu \mu}\right)\,,\nonumber\\
      C_{P} &= - N_2 \left(\frac{m_{\mu}}{M_{Z'}^2}\right) \left(a_{bs}g_{bs}+b_{bs}g_{bs}(q^2/m_{B}^2)\right) \left(g_R^{\mu \mu}- g_L^{\mu \mu}\right)\,,\nonumber\\
   C_{P}^{'} &= - N_2 \left(\frac{m_{\mu}}{M_{Z'}^2}\right) \left(a_{bs}g'_{bs}+b_{bs}g'_{bs}(q^2/m_{B}^2)\right) \left(g_R^{\mu \mu}- g_L^{\mu \mu}\right)\,,
 \end{align}
 where $N_2 = \sqrt{2} \pi/ (\alpha_{\rm em} G_F  V_{tb}V^*_{ts})\, \times 1/(q^2- M_{Z'}^2)$.  
 Here $C_9^{\rm NP}, C_{10}^{\rm NP},C_9^{\prime}$ and $C_{10}^{\prime}$  are WCs corresponding to NP VA operators defined in eq.~\eqref{heff-va} whereas the WCs $C_{P}$ and  $C_{P}^{'}$ correspond to
\begin{equation}
  \mathcal{H}^{\rm P} =
  -\frac{\alpha_{\rm em} G_F}{\sqrt{2} \pi} V_{ts}^* V_{tb} m_b
  \left[ C_{P} (\overline{s}  P_L b)
    (\overline{\mu} \gamma_{5} \mu)
    + C_{P}^{'} (\overline{s} P_R b)
    (\overline{\mu} \gamma_{5} \mu) \right].
\end{equation}

\begin{table*}[hbt]
\begin{center}
\begin{tabular}{|l|c|c|c|}
  \hline\hline
Scenario	 & Couplings & Wilson coefficients & Total $\Delta \chi^2$  \\
  \hline
ZL-I   & $F(q^2)\equiv 1,\,g_L^{\mu \mu}=g_R^{\mu \mu}=6.4 \times 10^{-4}$ & $C_{9}^{\rm NP} = N_2 \, g_{bs}g_{\mu \mu}$  &15.03\\ 
&$g_{bs}g_{\mu \mu}=7.04 \times 10^{-10}$, $g'_{bs}g_{\mu \mu}= 4.22 \times 10^{-10}$  & $C_{9}^{'} =N_2\, g'_{bs}\,g_{\mu \mu}$ &\\
\hline
ZL-II   & $a_{bs}\neq0,\,g_L^{\mu \mu}=g_R^{\mu \mu}$ & $C_{9}^{\rm NP} = N_2\,g_{bs}\,g_{\mu \mu}$ & 30.60\\ 
& $g_{bs}g_{\mu \mu}= 1.65 \times 10^{-9}$, $g'_{bs}g_{\mu \mu}=-4.21 \times 10^{-10}$ & $C_{9}^{'} = N_2\,g'_{bs}\,g_{\mu \mu}$ &\\ 
\hline
ZL-III  & $a_{bs}\neq0,\,g'_{bs}=0,\,g_L^{\mu \mu}\neq g_R^{\mu \mu}$ &  $C_{9}^{\rm NP} = N_2\, g_{bs}\, g_{\mu \mu}$  & 29.54\\ 
& $g_{bs}g_{\mu \mu}=1.54 \times 10^{-9}$, $g_{bs}g'_{\mu \mu}=1.56 \times 10^{-12}$ & $C_{10}^{\rm NP} = N_2\,g_{bs}\, g'_{\mu \mu}$ &\\
& & $C_P = -2 N_2\, \left(m_{\mu}/M_{Z'}^2 \right) \,g_{bs}\, g'_{\mu \mu}$ &\\
\hline
ZL-IV  & $a_{bs}\neq0,\,g_{bs}=0,\,g_L^{\mu \mu}\neq g_R^{\mu \mu}$ & $C_{9}^{'} =N_2\, g'_{bs}\, g_{\mu \mu}$  & 4.06\\ 
& $g'_{bs}g_{\mu \mu}=3.34 \times 10^{-10}$, $g'_{bs}g'_{\mu \mu}=1.16 \times 10^{-12}$ & $C_{10}^{'} = N_2\, g'_{bs}\, g'_{\mu \mu}$ &\\ 
& & $C_P^{'} = -2 N_2\, \left(m_{\mu}/M_{Z'}^2 \right) \, g'_{bs}\, g'_{\mu \mu}$ &\\
\hline
ZL-V      & $a_{bs}=0,\,g_L^{\mu \mu}=g_R^{\mu \mu}$  & $C_{9}^{\rm NP} = N_2\, (q^2/m^2_{B})\, g_{bs}  \,g_{\mu \mu}$  &49.54\\ 
& $g_{bs}g_{\mu \mu}=2.59 \times 10^{-8}$, $g'_{bs}g_{\mu \mu}=-9.38 \times 10^{-9}$ & $C_{9}^{'} = N_2\, (q^2/m^2_{B})\, g'_{bs} \, g_{\mu \mu}$ &\\ 
\hline
ZL-VI  & $a_{bs}=0,\, g'_{bs}=0,\, g_L^{\mu \mu} \neq g_R^{\mu \mu}$ & $C_{9}^{\rm NP} = N_2\, (q^2/m^2_{B})\, g_{bs} \, g_{\mu \mu}$ & 44.46 \\ 
& $g_{bs}g_{\mu \mu}=2.30 \times 10^{-8}$, $g_{bs}g'_{\mu \mu}=-0.56 \times 10^{-11}$ & $C_{10}^{\rm NP} =  N_2\, (q^2/m^2_{B})\, g_{bs}\, g'_{\mu \mu}$ &\\ 
& & $C_P = -2 N_2\, \left(m_{\mu}/M_{Z'}^2 \right) \,(q^2/m_{B}^2)\, g_{bs}\, g'_{\mu \mu}$ &\\
\hline
ZL-VII  & $a_{bs}=0,\, g_{bs}=0,\, g_L^{\mu \mu} \neq g_R^{\mu \mu}$ &  $C_{9}^{'} = N_2\, (q^2/m^2_{B})\, g'_{bs} \, g_{\mu \mu}$ & 3.66\\ 
& $g'_{bs}g_{\mu \mu}= 4.01 \times 10^{-9}$, $g'_{bs}g'_{\mu \mu}=9.64 \times 10^{-12}$ & $C_{10}^{'} = N_2\,(q^2/m^2_{B})\,g'_{bs} \, g'_{\mu \mu}$ &\\
& & $C_P^{'} =  -2 N_2\, \left(m_{\mu}/M_{Z'}^2 \right) \,(q^2/m_{B}^2)\, g'_{bs} \,g'_{\mu \mu}$ &\\ [2pt]
  \hline\hline
\end{tabular}
\caption{NP scenarios generated in light $Z'$ model. In ZL-II - ZL-IV, we define $a_{bs}g_{bs}\equiv g_{bs} $ and $a_{bs}g'_{bs}\equiv g'_{bs} $ whereas for ZL-V - ZL-VII, $b_{bs}g_{bs}\equiv g_{bs} $ and $b_{bs}g'_{bs}\equiv g'_{bs} $. The value of $\chi^2_{\rm SM}$ is  $\approx 197$.}
\label{scenarios}
\end{center}
\end{table*}

\begin{table*}[h!]
\centering
\begin{tabular}{|c|c|c|c|c|c|c|c|} \hline
Scenario & Total $ \Delta \chi^2$ & \multicolumn{5}{|c|}{Prediction}   \\ 
\hline 
& &  ${R_K}_{[1.1-6]}$  & ${R_{K^*}}_{[0.045-1.1]}$ & ${R_{K^*}}_{[1.1-6]}$ & ${P'_{5}}_{[4-6]}$ & $B(B_s^{\phi \mu})_{[1-6]} \times 10^{-8}$ \\ \hline
Exp (1$\sigma$) & -- & [0.80, 0.89] & [0.57, 0.80]  & [0.55, 0.77]   & [-0.55, -0.32] & [2.67, 3.09] \\ \hline
ZL-V & 49.54 &  [0.79, 0.89]   &  [0.86, 0.88]   & [0.75, 0.83]   &  [-0.57, -0.43]   &  [3.96, 4.4]  \\ \hline
ZL-VI & 44.46 & [0.75, 0.84] & [0.88, 0.90]  &  [0.82, 0.88]  &  [-0.59, -0.45]       &  [4.37, 4.71]   \\ \hline
\end{tabular}
\caption{Fit results for favoured scenarios  in light $Z'$ model.}
\label{tab:lzp2}
\end{table*}

Depending upon the choice of couplings, several NP scenarios can be generated. These are listed in Table \ref{scenarios}. For these scenarios, we use constraints coming from  several $b \to s \mu^+ \mu^-$ observables as used in our model independent global fit. For scenario ZL-I, $\Delta M_s$ is also included in the fit. Except scenario ZL-I, all other
scenarios correspond to $ F(q^2)\neq 1$ for which we do not include constraints from $\Delta M_s$  and $B_s \to \mu^+ \mu^-$ as $F(q^2)$ is unknown for $q^2 \sim m_B^2$. For $\Delta M_s$ , we use the theoretical expression given in eq.~\eqref{Zp} whereas $\Delta M_s^\text{\rm NP}$ is assumed to be as large as the SM uncertainty i.e., $\Delta M_s^\text{NP} = (0 \pm 1.2 )\, \rm {ps}^{-1}$ \cite{DiLuzio:2019jyq}. Further for ZL-I scenario, $g_L^{\mu \mu}=g_R^{\mu \mu}$ is fixed at $6.4 \times 10^{-4}$ which is 2$\sigma$ upper limit coming from the current measurement of the anomalous magnetic moment of muon \cite{Abi:2021gix}.

The best fit values of the couplings, along with the $\Delta \chi^2$ are presented in Table \ref{scenarios}. 
We find that the current $b \to s \mu^+ \mu^-$ data favours scenarios ZL-V and ZL-VI. The 1$\sigma$ range of some of the observables having tension with the SM are given in Table \ref{tab:lzp2} for these favoured scenarios. It is obvious that neither ZL-V nor ZL-VI can resolve $R_{K^*}$ anomaly in the very low-$q^2$ bin. However, like heavy $Z'$ model, these
scenarios are able to resolve the tension between the measurements and SM predictions of $R_{K^{(*)}}$ in [1.1-6.0] bin as well as $P'_5$ in [4.0-6.0] bin. For $B(B_s \to \phi\, \mu^+ \,\mu^-)$, the improvement is marginal. Therefore, we conclude that this model doesn't have any additional advantage over the heavy $Z'$ models in resolving the current $b \to s \mu^+ \mu^-$ anomalies.

In fig.~\ref{fig:psZprime1}, we show the $1\sigma$ ranges for some of the LFUV observables. It is apparent that both allowed scenarios predict $R_{\phi}<1$ as well as $R_{K^{(*)}}<1$ in the high-$q^2$ region. 
While ZL-VI scenario 
prefers higher values of $R_\phi$ and $R_{K^*}$ in the high $q^2$ bin, ZL-V scenario allows these observables to be less that 0.70 at 1$\sigma$ level.  
The allowed values of $Q_{4,5}$ in the [4, 6] bin 
cannot discriminate between the two favoured scenarios. 
However, a non-zero measured value of ${Q_4}_{[15,19]}$ would favour ZL-V scenario while
a non-zero value of 
${Q_5}_{[15-19]}$ would indicate ZL-VI. 

\begin{table*}[h!]
\centering
\begin{tabular}{|c|c|c|c|c|c|c|c|} \hline
Scenario & Total $ \Delta \chi^2$ & Couplings (best-fit) & \multicolumn{5}{|c|}{Prediction}   \\ 
\hline 
& & & ${R_K}_{[1.1-6]}$  & ${R_{K^*}}_{[0.045-1.1]}$ & ${R_{K^*}}_{[1.1-6]}$ & $B_{[1-6]}^{Kee}  \times 10^{7}$ & $B_{[0.09-1]}^{K^*ee}  \times 10^{7}$ \\ \hline
Exp (1$\sigma$) & -- & -- & [0.80, 0.89] & [0.57, 0.80]  & [0.55, 0.77]  & [1.38, 1.74] & [2.2, 4]  \\ \hline
ZL-V & 51.73 & $g_{bs}g_{\mu \mu}=2.39 \times 10^{-8}$  &[0.75, 0.92]   &  [0.71, 0.90]   & [0.45, 0.85] & [1.53, 2.16] & [1.38, 1.70]   \\ 
&&$g'_{bs}g_{\mu \mu}=-9.40 \times 10^{-9}$ &&&&& \\ 
&&$g_{bs}g_{ee}=-1.40 \times 10^{-8}$ &&&&& \\ 
&&$g'_{bs}g_{ee}=1.00 \times 10^{-8}$ &&&&& \\ \hline
ZL-VI & 45.21 & $g_{bs}g_{\mu \mu}=2.43 \times 10^{-8}$   & [0.72, 0.89] & [0.87, 0.92]  &  [0.78, 0.92]     & [1.46, 1.93] & [1.38, 1.44]\\
&&$g_{bs}g'_{\mu \mu}=1.51 \times 10^{-11}$ &&&&& \\ 
&&$g_{bs}g_{ee}=4.60 \times 10^{-9}$ &&&&& \\ 
&&$g_{bs}g'_{ee}=1.8 \times 10^{-10}$ &&&&& \\ \hline
\end{tabular}
\caption{Fit results for favoured scenarios  in light $Z'$ model with couplings both to muons and electrons. Here $B^{Kee} \equiv B(B \to K e^+ e^-)$ and $B^{K^*ee} \equiv B(B \to K^* e^+ e^-)$. The experimental values of  $B^{Kee}$ and $B^{K^*ee}$ are taken from refs. \cite{LHCb:2014vgu} and \cite{LHCb:2013pra}, respectively. }
\label{tab:lzp2-e}
\end{table*}

 We now consider a 25 MeV $Z'$ that couples both to muons and electrons.  It would be interesting to see whether such a model can provide a better resolution of $R_{K^*}$ anomaly in the very low-$q^2$ bin. In this model, we  consider NP scenarios ZL-V and Zl-VI which provided a good fit to $b \to s \mu^+ \mu^-$ data. Due to couplings with electrons, additional WCs are generated which  are obtained by replacing $\mu$ by $e$ in the  expression of WCs given in Table \ref{scenarios}. We redo the fit for scenarios ZL-V and Zl-VI. We do not include any additional observables in the fit. The NP WCs in $b \to s e^+ e^-$ sector are constrained by $R_{K^{(*)}}$. Instead, we predict the branching ratios of $B \to (K,\,K^*) e^+ e^-$ and compare them with their experimental values. The fit results are shown in Table \ref{tab:lzp2-e}.

It is evident from Table \ref{tab:lzp2-e} that the scenario ZL-V can now accommodate  the measurement of $R_{K^*}$ in the very low-$q^2$ bin whereas scenario ZL-VI can only provide a marginal improvement as compared to the couplings only to muons. We also find that the  predicted and the experimental value of the branching ratio of $B \to K e^+ e^-$ is consistent with each other whereas for $B \to K^* e^+ e^-$, the agreement is at $\sim$ 90\% C.L.

 Due to the fact that the $Z'$ couples to electron, it will decay on-shell into electron-positron pair with a branching ratio of $\sim 1$. Such a decay can lead to a finite contribution to the decay width of $Z'$ which is given as 
\begin{eqnarray}
\Gamma (Z' \to e^+ e^-) &=& \left[\frac{(g_L^{ee})^2+(g_R^{ee})^2}{12\pi}\right] M_{Z'}  \left[1-\frac{4m_e^2}{M_{Z'}^2} \right]^{1/2} \left[ 1+\frac{2m_e^2}{M_{Z'}^2}\right]\,.
\end{eqnarray}
Using the upper bound on $g_{ee}$ as obtained in \cite{Datta:2017ezo,NA482:2015wmo}, $\Gamma (Z' \to e^+ e^-)$ is predicted to be extremely small as compared to the invariant mass $q^2$ in $b \to s \ell^+ \ell^-$ transitions. Therefore, in the current scenario,  the inclusion of $\Gamma (Z' \to e^+ e^-)$ in the NP WCs are not expected to provide any observables effects. We check the validity of this statement by performing a fit for the scenario ZL-V after including $\Gamma (Z' \to e^+ e^-)$ in the NP WCs. Indeed, we find that the effects are negligible.

\begin{table}[h!]
\centering
\begin{tabular}{|c|c|c|} \hline
Observables & Experimental limits \cite{Grygier:2017tzo,Lees:2013kla,Lutz:2013ftz} & \multicolumn{1}{|c|}{Prediction}  \\ \hline
 &  &   ZL-V  \\ \hline
$B(B^0 \to K^0 \nu \bar{\nu})$ & $2.6 \times 10^{-5}$  & $3.9 \times 10^{-6}$     \\ \hline
$B(B^0 \to K^{*0} \nu \bar{\nu})$ & $1.8 \times 10^{-5}$ &  $9.1 \times 10^{-6}$    \\ \hline
$B(B^+ \to K^+ \nu \bar{\nu})$ & $1.6 \times 10^{-5}$  &  $4.2 \times 10^{-6}$    \\ \hline
$B(B^+ \to K^{*+} \nu \bar{\nu})$ & $4.0 \times 10^{-5}$ & $9.8 \times 10^{-6}$      \\ \hline
\end{tabular}
\caption{Predictions of branching ratios of decays induced by $b \to s \nu \bar{\nu}$ transitions in the light $Z'$ model with couplings both to muons and neutrinos. The experimental as well as theoretical limits are at  90\% C.L. The predictions for ZL-VI scenario are the same. }
\label{pred-bsnunu}
\end{table}

 We also consider a scenario where the light $Z'$, apart from coupling to muons,  can also couple to neutrinos. We assume that the couplings are driven by the $SU(2))_L$ symmetry. In this case the $Z'$ will decay on-shell into $\nu_{\mu}\bar{\nu_{\mu}}$ with a branching ratio of $\sim 1$. The decay width of $Z'$ is then given by
 \begin{equation}
 \Gamma (Z' \to \nu_{\mu}\bar{\nu_{\mu}}) = \left[\frac{(g_L^{\mu \mu})^2}{24 \pi}\right] M_{Z'}\,.
 \end{equation}
Using the constraints from the neutrino trident data,  $ \Gamma (Z' \to \nu_{\mu}\bar{\nu_{\mu}}) $ is predicted to be too small to make any observable effects in the fits. We have verified  this for the scenario ZL-V.  Moreover, $b \to s \nu \bar{\nu}$ transition can also be generated in this model. The effective Hamiltonian for $b \to s \nu \bar{\nu}$ transition is given by \cite{Buras:2014fpa}
\begin{eqnarray}
{\mathcal H}_{b \to s \nu \bar{\nu}} = - \frac{\sqrt{2}\alpha G_F}{\pi} V_{tb}V^*_{ts} \sum_{\ell} C_L^\ell \left(\bar{s}\gamma^{\mu}P_L b\right) \left(\bar{\nu_\ell}\gamma_{\mu}P_L \nu_\ell\right),
\end{eqnarray}
where $C_L^\ell=C_L^{\rm SM}+C^{\ell \ell}_{\nu}{(\rm NP)}$. The NP contribution $C^{\ell \ell}_{\nu}(\rm NP)$ in the light $Z'$ model is given by
\begin{equation}
C^{\mu \mu}_{\nu}({\rm NP}) = - \frac{\pi}{\sqrt{2}\,\alpha G_F\,  V_{tb}V^*_{ts}} \frac{F(q^2)\,g_{bs}\,g_L^{\mu \mu} }{q^2 - M^2_{Z'}}\,.
\end{equation}
The SM WC is $C_L^{\rm SM}=-X_t/ \sin^2 \theta_W$, where $X_t=1.461 \pm 0.017$. Using the fit results for working scenarios ZL-V and ZL-VI, we obtain predictions of decays induced by the quark level transition $b \to s \nu \bar{\nu}$. The predicted values are shown in Table \ref{pred-bsnunu}. It can be seen that the predicted values are well within the current experimental limits. 

The measurements of muon $g-2$ can provide constraints on the $Z'$ couplings to muons. The contribution to muon $g-2$ in the light $Z'$ model is obtained as \cite{Buras:2021btx,Crivellin:2022obd}
\begin{equation}
\Delta a_{\mu} = \frac{m_{\mu}^2}{12\pi^2 M_{Z'}^2} {\rm Re} \left[(g_{\mu \mu})^2-(g'_{\mu\mu})^2\right],
\end{equation} 
with $\Delta a_{\mu}=a_{\mu}^{\rm exp} - a_{\mu}^{\rm SM} = (251\pm59) \times 10^{-11}$ \cite{Muong-2:2006rrc,Muong-2:2021ojo,Aoyama:2020ynm}. The observable $\Delta a_{\mu}$ is expected to provide constraints  on the muon couplings $g_{\mu \mu}$ and $g'_{\mu \mu}$. However due to  the fact that $g_{bs} $ and $g'_{bs} $ couplings are unconstrained in the context of the considered model for scenarios ZL-V and ZL-VI, the products of these quarks and muons couplings which appear in $b \to s \mu^+ \mu^-$ transition is likely to remain unchanged. As our predictions in Table \ref{tab:lzp2} depend upon the products of $Z'$ couplings to quarks and leptons, the results are expected to remain unchanged.  By performing a three parameter fit,  ($g_{bs}$, $g'_{bs}$, $g_{\mu\mu}$) for scenario ZL-V and ($g_{bs}$, $g'_{\mu\mu}$, $g_{\mu\mu}$) for ZL-VI, to $b \to s \ell \ell$ data and then to a combination of  $\Delta a_{\mu}$ and $b \to s \ell \ell$ data set, we find that the results obtained in Table \ref{tab:lzp2} remains unaltered.

 \section{Conclusions}
\label{concl}
In this work, we provide an update of the model independent global analyses of $b \to s \ell \ell$ data in the light of the recent measurements of LFUV observables $R_{K_S^0}$  and $R_{K^{*+}}$, along with the updated measurements of the branching ratio and angular  observables in $B_s \to \phi \mu^+ \mu^-$ decay  by the LHCb collaboration. Assuming NP only in the muon sector and considering either one or two independent NP operators at a time, the NP solutions are determined on the basis of $\Delta \chi^2$ which is the difference between the $\chi^2$ in the SM and the NP scenario. In addition, we also predict LFUV observable $R_{\phi}$, $R_{K^{(*)}}$ in the high $q^2$ bin, as well as $Q_{4,5}$ observables. We find that
\begin{itemize}
\item for 1D scenarios,  $C_9^{\rm NP}$ and $C_{9}^{\rm NP}=-C_{10}^{\rm NP}$ provide a good fit fit to the data with $\Delta \chi^2 \approx 45$. However, $\Delta \chi^2$ for  $C_{9}^{\rm NP}=-C'_{9}$ scenario slips to $\approx 30$ which is marginally below the value for  the $C_{10}^{\rm NP}$ scenario. Hence the updated data now prefers  $C_{10}^{\rm NP}$ scenario over $C_{9}^{\rm NP}=-C'_{9}$. This is mainly due to the fact that the later fails to accommodate the  measurements of $R_K$ and $R_{K_S^0}$. 

\item at the best fit point, all four 1D solutions predict $R_{\phi} < 1$ in both low and high-$q^2$ bins and $R_{K^*}< 1$ in the high-$q^2$ bin. The value of $R_{K}$ in the high-$q^2$ bin
is predicted to be less than unity for all 1D scenarios except $C_{9}^{\rm NP}=-C'_{9}$ for which the value is $\approx 1$. This is the only 1D solution which can invoke the 1$\sigma$ lower limit of $R_{\phi}$ and $R_{K^*}$ in the high-$q^2$ bin to be less than 0.75.
Moreover,  a precise measurement of magnitude and sign of $Q_{4,5}$ can provide an unique identification of  $C_9^{\rm NP}$, $C_{10}^{\rm NP}$ and $C_{9}^{\rm NP}=-C'_{9}$ solutions.

\item the 2D scenarios  $(C_9^{\rm NP}, C_{10}^{\prime} )$, $(C_9^{\rm NP}, C_{9}^{\prime} )$ and $(C_9^{\rm NP}, C_{10}^{\rm NP} )$ are still favoured with $\Delta \chi^2 \approx$ 51, 48 and 48, respectively.

\item all 2D scenarios predict $R_{\phi} < 1$ in both low and high-$q^2$ bins, $R_{K^{(*)}}<1$ in the high $q^2$ bin 
and allow ${Q_5}_{[4-6]}>0.25$. Interestingly, a non-zero measurement of ${Q_4}_{[15-19]}$ would disfavour $(C_9^{\rm NP}, C_{10}^{\rm NP} )$  over other two scenarios. 
 
\end{itemize}

We then consider a generic TeV scale $Z'$ model which generates the 2D favored solutions $(C_9^{\rm NP}, C_{9}^{\prime} )$ and $(C_9^{\rm NP}, C_{10}^{\rm NP} )$ as well as 1D ones $C_9^{\rm NP} $ and 
$C_9^{\rm NP} = -C_{10}^{\rm NP}$. In these models there are additional constraints coming from  $B_s -\bar{B_s}$ mixing and neutrino trident data. We find that
$\Delta \chi^2 $ is approximately the same, $ (42-47)$,  for all four models indicating that they are viable models to explain the $b \to s \mu^+ \mu^-$ anomalies.  However none of these models can resolve the tension between  the SM and measured value of $R_{K^*}$ in  the very low-$q^2$ bin as well as the branching ratio of $B_s \to \phi \mu^+ \mu^-$ in the low-$q^2$ bin. Moreover, a precise measurement of $Q_4$ and $Q_5$ in the high-$q^2$ bin can not only discriminate amongst the two 1D and 2D models but can also disentangle 1D and 2D $Z'$ framework. 

We finally consider a model with 25 MeV  $Z'$ having a $q^2$  dependent $b - s$ coupling along with coupling only to muons. This model generates several 2D scenarios. We find that  the  scenarios which induce  $({\cal O}_9,\,{\cal O}_{10})$ and $({\cal O}_9,\,{\cal O}'_{9})$  NP operators provide a good fit to all  $b \to s \mu^+ \mu^-$ data. However, this model doesn't have any additional advantage over the heavy $Z'$ models in resolving the current $b \to s \mu^+ \mu^-$ anomalies. The $Q_4$ observable in the high-$q^2$ bin can be a good discriminant between the two favoured scenarios. If the coupling to electron is also allowed then the favored scenario which generates   $({\cal O}_9,\,{\cal O}_{10})$ operators can  accommodate  the measurement of $R_{K^*}$ in the very low-$q^2$ bin whereas the other scenario can only provide a marginal improvement as compared to the couplings only to muons.

\bigskip
\noindent
{\bf Acknowledgements}: The work of A.K.A. is supported by SERB-India Grant CRG/2020/004576. SG acknowledges 
support from the ANR under contract n. 202650 (PRC `GammaRare').

\end{document}